\documentclass[aps,prb,amsfonts,amssymb,longbibliography,onecolumn,amsmath,floatfix,showpacs]{revtex4-2}

\usepackage{amsbsy}
\usepackage{graphicx} 
\usepackage{subcaption}
\usepackage{bm} 
\usepackage[T2A]{fontenc}
\usepackage[cp1251]{inputenc}
\usepackage[english]{babel}
\usepackage[capitalise]{cleveref}

\newcommand{\be}{\begin{equation}}
\newcommand{\ee}{\end{equation}}
\def\bal#1\eal{\begin{align}#1\end{align}}
\def\bml#1\eml{\begin{multline}#1\end{multline}}
\def\bald#1\eald{\begin{aligned}#1\end{aligned}}
\newcommand{\bea}{\begin{eqnarray}}
\newcommand{\eea}{\end{eqnarray}}

\DeclareMathOperator{\sgn}{sgn}

\begin{document}

\title{Boundary conditions for the order parameter and the proximity influenced internal phase differences in double 
superconducting junctions}

\author{Yu.\,S.~Barash}

\affiliation{Institute of Solid State Physics of the Russian Academy of Sciences,
Chernogolovka, Moscow District, 2 Academician Ossipyan Street, 142432 Russia}

\date{\today}

\begin{abstract}
This paper gives an overview of the unconventional dependence of internal phase differences on the external phase
difference in superconductor-normal metal-superconductor (\mbox{SINIS}) and superconductor-superconductor-superconductor
(\mbox{SISIS}) tunnel double junctions. The results are obtained within the Ginzburg-Landau (GL) approach that includes
boundary conditions for the superconductor order parameter in the presence of a Josephson coupling through
interfaces. The boundary conditions are derived within the GL theory and substantiated microscopically. The absence of
the one-to-one correspondence between external $\phi$ and internal $\chi_{1,2}$ phase differences in the junctions is
shown to occur in two qualitatively different ways, both of which result in the range of $\chi_{1,2}$ reduced and
prevent the $4\pi$-periodic current-phase relation $j=j_c\sin\frac{\phi}{2}$. In \mbox{SINIS} junctions, the effect of
the supercurrent-induced phase incursion $\varphi$ between the end faces of the central electrode of mesoscopic length 
$L$ can play a crucial role. In \mbox{SISIS} junctions, there occurs the regime of interchanging modes, which is modified as $L$
decreases. The proximity and pair breaking effects in the double junctions with closely spaced interfaces are addressed.
\end{abstract}

\maketitle

\section{Introduction}

A phase dependent Josephson coupling of two superconductors linked to each other by quasiparticle tunneling through an 
interlayer, gives rise to the Josephson current through the system~\cite{Josephson1962,Josephson1964,Josephson1965}.
If two Josephson junctions, connected in series in a \mbox{SISIS} heterostructure, are at distance $L$ that far exceeds 
the coherence length $\xi$, the coupling between them is negligible in the absence of magnetic effects. However, under 
the opposite condition $L\alt\xi$, the proximity and interface pair breaking can have a strong effect on the transport 
processes, including the Josephson current. Static and dynamic interactions of two closely spaced Josephson junctions
can manifest in a variety of properties of mesoscopic superconducting heterostructures and play an important role in 
superconducting electronics~\cite{Kao1977_2,HansenLindelof1984,Likharev1984,Blackburn1987,Smith1990,Blamire1994,%
Kupriyanov1999,Goldobin1999,Nevirkovets1999,Golubov2000,Brinkman2001,Ishikawa2001,Brinkman2003,Blamire2006,Luczka2012,%
Linder2017,Barash2018,Bakurskiy2019}.

When a normal metal is placed between the superconductors, the Josephson coupling emerges as a consequence of 
superconducting correlations generated by proximity effects in the normal metal region~\cite{deGennes1964,Likharev1979,Belzig1999,%
Klapwijk2004,Golubov2004}. A number of hybrid systems with proximity-induced Josephson coupling 
through normal metal electrodes, have recently been the focus of scientific scrutiny~\cite{Petrashov1995,%
Devoret1996,Balashov1998,Schoen2001,Belzig2002,Esteve2008,Giazotto2010,Giazotto2011_2,Giazotto2011,Giazotto2014,%
Giazotto2015,Giazotto2017,Ryazanov2017,Barash2019,Kupriyanov2021}.

The phase difference $\phi$ between external superconducting electrodes in symmetric superconducting double junctions 
incorporates the internal phase differences $\chi_{1,2}$ across the interfaces of constituent \mbox{SIS} or \mbox{SIN}
junctions and the supercurrent-induced phase incursion $\varphi$ between the end faces of the central electrode of 
length $L$: $\phi=\chi_{1}+\chi_{2}+\varphi$. Both the external and internal phase differences can be identified 
experimentally and controlled, although only one of them is usually an independent variable in the equilibrium state. 
Thus the internal phase difference can be controlled by the magnetic flux through a superconducting ring involving only
one of two constituent contacts of a double junction. 

However, the problem of internal phase differences has not yet received sufficient attention in the literature, 
whereas various other properties of \mbox{SINIS} and \mbox{SISIS} superconducting double junctions have been thoroughly 
developed and applied to a wide range of heterostructural parameters. One of the reasons for this is that the relation
between the phase drops $\chi_{1,2}$ across the interfaces and the phase incursion $\varphi$ over the central lead at a 
given $\phi$ is usually simplified, assuming negligible values of either $\varphi$, or $\chi_{1,2}$. For 
example, the more advanced early attempts to describe the \mbox{SNS} junctions within the Ginzburg-Landau (GL) 
approach~\cite{VolkovA1971,Fink1976} considered the phase incursion and fully transparent interfaces, but used the 
specific boundary conditions and disregarded the phase drops. In particular, the boundary conditions (7) in 
Ref.~\cite{VolkovA1971}, correct under the conditions studied, do not apply to the systems with the Josephson coupling. 
The boundary condition (5) in Ref.~\cite{Fink1976} is incorrect in general. On the other hand, in microscopic theories 
of the superconducting double junctions a negligible current-induced phase incursion is usually assumed, as distinct 
from the phase drops~\cite{Kupriyanov1988,Kupriyanov1999,Golubov2000,Golubov2004}. Although the latter point is applied 
to \mbox{SISIS} junctions in a wide range of realistic parameters, the range gets narrower in \mbox{SINIS} junctions, 
where both $\chi_{1,2}$ and $\varphi$ should, generally speaking, be taken into account for describing the transport at 
mesoscopic values of $L$~\cite{Barash2019}.

Another aspect of the problem is that a fully self-consistent description should include the spatially dependent order 
parameter absolute value and its spatially dependent phase. This substantially complicates numerical simulations, as 
well as analytical calculations, since the phase changes incorporate sharp drops across thin interfaces and the phase 
incursion along the central electrode. An alternative approximate approach disregards the joint self-consistency 
conditions and assumes the order parameter phase to be either spatially constant, or a linear function of coordinates 
inside each of the individual electrodes. The latter approach substantially simplifies the solution and, in a variety 
of cases, can be justified. However, an implicit assumption of such a standard approximation is the one-to-one 
correspondence between the external $\phi$ and internal $\chi_{1,2}$ phase differences, which recently has been 
demonstrated to be invalid under certain conditions in the double and multiple 
junctions~\cite{Linder2017,Barash2018,Barash2019}. The one-to-one correspondence gets broken in the \mbox{SINIS} and 
\mbox{SISIS} superconducting double junctions under two qualitatively different scenarios characteristic of the behavior
of internal phase differences driven by the external phase difference and influenced by the proximity and pair breaking 
effects. It is one of the consequences of the broken correspondence that the $4\pi$-periodic current-phase relation 
$j=j_c\sin\frac{\phi}{2}$, which is based on the simple equalities $\chi_{1,2}=\chi$, $j=j_c\sin\chi$, $\phi=2\chi$, 
does not hold in the double tunnel junctions over the entire range of $\phi$ even when the phase incursion and proximity
effects are negligibly small.

This paper gives an overview of the results relating to the internal phase differences in symmetric \mbox{SINIS} and
\mbox{SISIS} double junctions, obtained within the GL approach involving the boundary conditions for the order parameter
that take into account the Josephson coupling through interfaces~\cite{Barash2018,Barash2019}. The problem of the 
corresponding boundary conditions will be discussed in detail below. Representing the self-consistency equation for the
complex order parameter, the GL equations and boundary conditions offer the simplest way for describing the spatial 
distribution of both the order parameter's absolute value and its phase in superconducting heterostructures. 

It is worth noting that if, in a problem, the specific GL temperature dependence of the quantities (i.e., their 
power-law dependence on $|T-T_c|$) is irrelevant, then as a rule, the corresponding results remain applicable on the
semiquantitative level to a substantially wider temperature range compared to the GL temperature region near 
$T_c$~\cite{JGEHarris2016}. This is expected to be the case in studying the internal phase differences and current-phase
relations in both types of superconducting double junctions considered below within the GL theory.

The paper is organized in the following way: the Josephson coupling and the boundary conditions for the order parameter 
are both described within the GL theory and microscopically derived in Sec.~\ref{sec: bc}; the phase relations in 
symmetric \mbox{SINIS} and \mbox{SISIS} superconducting double junctions are considered, based on the approach
developed, in Secs.~\ref{sec: sinis} and~\ref{sec: sisis} respectively.

\section{Josephson Coupling and Boundary Conditions in the GL Theory}
\label{sec: bc}

\subsection{Symmetric Junctions between Identical Superconductors}
\label{subsec: symm}

The Josephson effect was successfully described within the GL theory first for the case of a superconducting point 
contact involving a fully transparent small constriction~\cite{AslamazovLarkin1969}. The focus of this paper is another
standard type of superconducting junctions, where superconducting leads with identical constant cross sections are 
linked to each other by the quasiparticle transmission through planar interlayers with interfacial barriers. The 
Josephson part of the free energy functional near $T_c$, considered for the junction with a single thin 
interlayer, is the basis for getting the GL equations and boundary conditions at 
interfaces~\cite{deGennes1966,Yip1990,Sigrist1991,Lifshitz1995,Mineev1999}.  

The phase dependent part of the Josephson interface free energy per unit area has a well-known bilinear form
\be
{\cal F}_{\chi}^J=-g_J\bigl(\Psi_+\Psi_-^*+\Psi_-\Psi_+^*\bigr)=-2g_J\left|\Psi_-\right|\left|\Psi_+\right|\cos\chi,
\label{Jos1}
\ee
where the quantities $\Psi_{\pm}=|\Psi_\pm|e^{\alpha_\pm}$ are the order parameter complex values on opposite interface 
sides, $g_J$ is the Josephson coupling constant and $\chi=\alpha_--\alpha_+$. Hereafter, the thickness of a homogeneous
interlayer is assumed to be comparatively small for defining it to be zero within the GL theory. If the interlayer is 
placed at $X=X_{\text{int}}$, then $X_\pm=X_{\text{int}}\pm0$. This does not exclude, for example, metallic interlayers 
with thicknesses on the order of the Cooper pair size. 

When Eq. \eqref{Jos1} is considered in the tunneling limit, the quantities $|\Psi_\pm|$ are taken in the zeroth-order 
approximation in powers of the Josephson coupling constant and are therefore independent of the phase difference $\chi$. 
In this case one gets from \eqref{Jos1} the standard textbook expression for the Josephson supercurrent density. In 
terms of the chosen notations 
\be
j=\dfrac{2|e|}{\hbar}\dfrac{d{\cal F}_{\chi}^J}{d\chi}=
\dfrac{4|e|g_J}{\hbar}\left|\Psi_-\right|\left|\Psi_+\right|\sin\chi.
\label{Jos2}
\ee
For $g_J>0$ this is a $0$ junction, while one gets $\pi$ junction for $g_J<0$.

Although the phase-dependent interface free energy \eqref{Jos1} incorporates important basic features of the Josephson 
effect, it generally disagrees with the microscopic theory since it does not take into account the accompanying 
proximity effects of the Josephson origin. The disadvantage originates from the incorrect condition, at which the 
Josephson free energy part vanishes. Thus, assuming for identical superconducting leads equal order parameter values on 
opposite superconducting banks $\left|\Psi_\pm\right|=|\Psi|$ the Josephson junction energy is known to be proportional
to $|\Psi|^2(1-\cos\chi)$ and to vanish at $\chi=0$~\cite{Tinkham1996,Barone1982}. The latter expression must be 
generalized in the GL theory to address the free energy variation due to independently changing order parameter
values on opposite interface sides. The corresponding invariant term in the simplest case is
\be
{\cal F}^{J}_s=g_J\left|\Psi_+-\Psi_-\right|^2 =
g_J\Bigl[\left|\Psi_+\right|^2+\left|\Psi_-\right|^2-
\bigl(\Psi_+\Psi_-^*+\Psi_-\Psi_+^*\bigr)\Bigr].
\label{Jos3}
\ee

The expression \eqref{Jos3} implies the interlayer to be in the time-reversal invariant state and also symmetric with 
respect to the normal-axis inversion. 

While the Josephson contribution \eqref{Jos3} becomes proportional to $1-\cos\chi$ and vanishes at zero phase difference
$\chi=0$ under the condition $\left|\Psi_+\right|=\left|\Psi_-\right|$, it does not vanish when 
$\left|\Psi_+\right|\ne\left|\Psi_-\right|$:
\be
{\cal F}^{J}_{\chi=0}=g_J\bigl(\left|\Psi_+\right|-\left|\Psi_-\right|\bigr)^2.
\label{Jos4}
\ee

If $g_J>0$, the quantity ${\cal F}^{J}_{\chi=0}$ increases with a difference between the order parameter values 
$\left|\Psi_\pm\right|$. The associated proximity effect, that influences the near-interface order-parameter 
structure, can substantially differ from what would follow from \eqref{Jos1}.

The superconductivity can also be weakened (or enhanced) by its proximity to the boundary with an adjacent material, 
irrespective of the presence or absence of the supercurrent depairing and of the Josephson coupling over the boundaries.
The surface depairing is known to occur, for example, near surfaces with normal metals and/or 
magnets~\cite{deGennes1964,ROZaitsev1965,deGennes1966,Ivanov1981,Sauls1988,Nazarov2009}. Possible reasons for the 
surface superconductivity enhancement have been considered in Refs.~\cite{Babaev2020,CroitoruShanenko2020,Babaev2021} 
and references therein. The depairing effects in the near-surface region can also be associated with an anisotropic 
pairing structure in unconventional superconductors~\cite{Buchholtz1981,Barash1995,Shiba1995a,Shiba1995b,Shiba1995c,%
Shiba1996,Nagai1995,Sauls1995a,Sauls1995b,Alber1996,Agterberg1997}.

In the GL free energy with a single-component order parameter, terms of the form $g|\Psi_\pm|^2$ are 
responsible for the surface (interface) pair breaking at $g>0$, or for enhancing the condensate density near
the surface (interface) at $g<0$. After adding the terms to the Josephson contribution 
\eqref{Jos3}, one gets the following quadratic form for the interface GL free energy per unit area
\be
{\cal F}^{int}=(g+g_J)\bigl(\left|\Psi_+\right|^2+\left|\Psi_-\right|^2\bigr)
-g_J\bigl(\Psi_+\Psi_-^*+\Psi_-\Psi_+^*\bigr).
\label{Jos5}
\ee

As distinct from the bulk GL free energy, no terms involving the order parameter derivatives or its higher powers are 
usually required in the interface contributions \eqref{Jos3} and \eqref{Jos5}, since neither of the corresponding 
coefficients vanishes there in view of the presumed absence of a zero-field quasi-two-dimensional phase transition in 
thin interfaces in question. Exclusions may concern small interface terms, not considered here, that result in 
qualitatively new kind of effect.

As seen in \eqref{Jos5}, the Josephson coupling constant $g_J$ enters not only the phase-dependent expression \eqref{Jos1}, but also 
the pair breaking terms, where $g_J$ is combined with $g$. This influences the near-surface spatial structure of the GL 
order parameter on the interface sides and contrasts to the standard analysis of the Josephson effect and boundary 
conditions within the GL theory, where the Josephson coupling was considered to be irrelevant to the surface pair 
breaking terms $\propto|\Psi_\pm|^2$ in the free energy. Therefore no such terms were assumed to appear in default of 
the surface pair breaking constant $g$, for example, in the case of conventional dielectric surfaces~\cite{deGennes1966}.

The boundary conditions are obtained in the GL theory from the vanishing free energy variation over the order parameter 
taken at a boundary or an interface side. As is well known, in addition to a surface or interface free energy, the 
bulk terms containing order parameter derivatives also contribute to the boundary conditions. Thus, in the simplest case
discussed, the gradient bulk term $K\bigl|d\Psi(x)/dx\bigl|^2$ leads after the variation over $\Psi^*(x)$ not only to 
the second-order derivative $-Kd^2\Psi/dx^2$ in the GL equations for each of the electrodes, but also in the boundary 
terms $Kd\Psi/dx\bigl|_{x_2}$ and $-Kd\Psi/dx\bigl|_{x_1}$ that, respectively, enter the boundary conditions at the 
upper and the lower limits of the integration. Based on such a prescription and using \eqref{Jos5}, one gets the 
following boundary conditions for the order parameter on opposite sides of the interface
\be
K\left(\dfrac{d\Psi}{dX}\right)_\pm=\pm(g+g_J)\Psi_\pm \mp g_J\Psi_\mp.
\label{Jos6}
\ee

The contributions from the phase dependent Josephson coupling $\mp g_J\Psi_\mp$, from the surface pair 
breaking $\pm g\Psi_\pm$ and the interface proximity effects of the Josephson origin $\pm g_J\Psi_\pm$ have been 
taken into account on the right-hand side of the condition \eqref{Jos6}. The use of \eqref{Jos1} instead of 
\eqref{Jos5} would ignore both the surface pair breaking and the term $\pm g_J\Psi_\pm$, which 
can substantially distort the boundary condition. 

When $g=0$, \eqref{Jos6} is reduced to
\be
K\left(\dfrac{d\Psi}{dX}\right)_\pm=g_J\left(\Psi_+-\Psi_-\right).
\label{Jos8}
\ee
Here the derivative of the order parameter, taken on one side of the interface with the corresponding coefficient, is 
linked by the Josephson coupling strength with the difference between the order parameter values on the two 
interface sides.

The boundary conditions \eqref{Jos6} have been used for describing single and double Josephson junctions within the GL 
theory~\cite{Barash2012,Barash2012_2,Barash2012_3,Barash2014_2,Barash2014_3,Barash2017,Barash2018,Barash2019}.

\subsection{Junctions with the Broken Symmetries}
\label{subsec: asymjunc}

One might think that the Josephson coupling-induced free energy \eqref{Jos3} applies also to a junction with 
electrodes made of different superconducting materials. This is generally not the case as is evidenced by the fact that
changing the material dependent normalization of the order parameter in \eqref{Jos3} results not only in a 
re-defining of the coefficient $g_J$, but also in an additional independent parameter associated with the difference 
between normalization factors on the opposite sides of the interface. A well-known change of the normalization factor is
associated with switching from the gap function to the standard GL order 
parameter~\cite{Gor'kov1959ar,Gor'kov1959br,Abrikosov1988}. Although a joint consideration of different superconducting
materials within the GL theory is known to be seriously restricted by the requirement that their critical temperatures 
$T_{c\pm}$ must be close to each other, various other independent material parameters can differ substantially. This 
concerns both \mbox{SIS'} junctions at $T<T_{c\pm}$ and \mbox{SIN} junctions at $T_{c-}<T<T_{c+}$~\cite{Abrikosov1988}.

Therefore, one should write a generalized version of \eqref{Jos3}:
\be
{\cal F}^{J}=\left|\beta_{+}\Psi_+-\beta_{-}\Psi_-\right|^2.
\label{Jos10a}
\ee
Here $\beta_{\pm}$ are real coefficients that can have identical or opposite signs.

Introducing $g_{J}=\beta_+\beta_-$ and $g_{J\pm}=|\beta_\pm|^2$, one obtains from \eqref{Jos10a}
\be
{\cal F}^{J}=g_J\left|\Psi_+-\Psi_-\right|^2+(g_{J+}-g_J)\left|\Psi_+\right|^2 
+(g_{J-}-g_J)\left|\Psi_-\right|^2.
\label{Jos11}
\ee

Unlike \eqref{Jos3}, a change of the order parameter normalization in \eqref{Jos10a} or \eqref{Jos11} does not result in
new terms, but only in the re-defining of the coefficients $\beta_{\pm}$, $g_J$ and $g_{J\pm}$. 

The expression \eqref{Jos11} also applies to the case of an interlayer with the broken normal-axis inversion. After 
adding to \eqref{Jos11} the surface pair-breaking terms $g_\pm^{\text{surf}}|\Psi_\pm|^2$ and introducing notations
$g_\pm=g_\pm^{\text{surf}}+(g_{J\pm}-g_J)$, one gets for the interface free energy
\be
{\cal F}^{int}=(g_++g_{J})\left|\Psi_+\right|^2+(g_-+g_{J})\left|\Psi_-\right|^2 
-g_J\bigl(\Psi_+\Psi_-^*+\Psi_-\Psi_+^*\bigr).
\label{Jos12}
\ee

The boundary conditions that follow from \eqref{Jos12} are
\begin{align}
&K_+\left(\dfrac{d\Psi}{dX}\right)_+=\bigl(g_++g_{J}\bigr)\Psi_+ -g_J\Psi_-,
\label{Jos6a}\\
&K_-\left(\dfrac{d\Psi}{dX}\right)_-=g_J\Psi_+-\bigl(g_-+g_{J}\bigr)\Psi_-.
\label{Jos7a}
\end{align}

The free energy \eqref{Jos12} and the boundary conditions \eqref{Jos6a}, \eqref{Jos7a} are similar to \eqref{Jos5} and 
\eqref{Jos6}, respectively. The difference is associated with distinct coefficients $g_\pm$ of the pair breaking terms 
on opposite interface sides compared with a common coefficient $g$ in \eqref{Jos5} and \eqref{Jos6} in symmetric 
junctions. In the absence of surface contributions $g_\pm^{\text{surf}}=0$, the quantities $g_{\pm}$ do not vanish here
due to the finite differences $g_{J\pm}-g_{J}$ of the Josephson origin.

It is worth noting that in a more involved case of chiral interlayers with the broken time reversal symmetry, the 
coefficients in \eqref{Jos10a} can take complex values $\beta_{\pm}=|\beta_{\pm}|e^{\gamma_{\pm}}$ with different phases
$\gamma_\pm$. The corresponding interface free energy is
\be
{\cal F}^{int}=(g_{J+}+g_+)\left|\Psi_+\right|^2+(g_{J-}+g_-)\left|\Psi_-\right|^2 
-2g_J|\Psi_-\Psi_+|\cos(\chi-\chi_0),
\label{Jos10}
\ee
where $g_{J\pm}=|\beta_\pm|^2$, $g_J=|\beta_+\beta_-|$, $\chi_0=\gamma_+-\gamma_-$.

The associated boundary conditions are
\begin{align}
&K_+\left(\dfrac{d\Psi}{dX}\right)_+=\bigl(g_++g_{J+}\bigr)\Psi_+-g_Je^{-i\chi_0}\Psi_-,
\label{Jos6b}\\
&K_-\left(\dfrac{d\Psi}{dX}\right)_-=g_Je^{i\chi_0}\Psi_+-\bigl(g_-+g_{J-}\bigr)\Psi_-.
\label{Jos7b}
\end{align}

The anomalous Josephson current, which is proportional to $\sin(\chi-\chi_0)$ in the tunneling limit, follows from 
\eqref{Jos10} in agreement with what is known for such a 
case~\cite{Buzdin2008,Zazunovetal2009,LiuChan2010,Kouwenhoven2016,Silaevetal2017}. A joint effect of the interfacial 
spin-orbit coupling and exchange field can also induce the spontaneous boundary currents~\cite{MironovBuzdin2017}, 
described by introducing an additional specific first-order gradient term in the GL interface free 
energy~\cite{Edelstein1996,Samokhin2004,Kaur2005,Edelstein2021}. Since both the latter effect and the anomalous 
Josephson effect itself are beyond the scope of this paper, complex values of $\beta_{\pm}$ will not be considered 
below.

\subsection{The Josephson Coupling in the GL Theory: a Microscopic Viewpoint}
\label{subsec: micro}

Quasiclassical theory of dirty superconductors probably offers the simplest way for a microscopic derivation of the 
boundary conditions for the GL order parameter. Thus the Kupriyanov-Lukichev boundary conditions, which are known to be 
applied within the Usadel approach to the Green functions at tunnel interfaces~\cite{Kupriyanov1988}, can be linearized 
in the anomalous Green function $F$ as it becomes small near $T_c$. Focusing on symmetric tunnel junctions with 
identical superconducting electrodes one obtains the following linearized relationship
\be
\dfrac{\sigma}{G_N}\left(\dfrac{dF}{dX}\right)_\pm=F_+-F_- .
\label{Jos14}
\ee
Here $G_N$ is the interface conductance per unit area and $\sigma$ is the normal-state conductivity of the
superconducting leads.

Similarly to Subsec.~\ref{subsec: symm}, it is presumed in \eqref{Jos14} that the first-order terms dominate the 
boundary conditions near $T_c$. Since the anomalous Green function $F$ and the gap function $\Delta$ are both small near
$T_c$, and their spatial derivatives introduce additional small factors, the relation $\Delta=\omega F$, where $\omega$
is the Matsubara frequency, follows, in the first approximation, from the Usadel equations. Therefore, the boundary 
condition for the gap function does not change as compared to \eqref{Jos14}:
\be
\dfrac{\sigma}{G_N}\left(\dfrac{d\Delta}{dX}\right)_\pm=\Delta_+-\Delta_- .
\label{Jos15}
\ee
The same result can also be obtained from \eqref{Jos14} based on the self-consistency condition
\be
\Delta(x)=\pi\lambda T\sum_{|\omega|<\omega_D}F(x,\omega).
\label{Jos15a}
\ee 
Here the BCS coupling constant $\lambda$ and the Debye frequency $\omega_D$ of the superconducting electrodes enter
the equality.

Eq.~\eqref{Jos15} has a form of the GL boundary conditions \eqref{Jos8} applied to symmetric junctions with no surface 
pair breaking, i.e., at $g=0$. The gap function near $T_c$ is known to play the role of the order parameter and to 
differ from the standard  GL order parameter $\Psi$ only by the normalization factor and the associated modified 
microscopic definitions of GL coefficients~\cite{Gor'kov1959ar,Gor'kov1959br,Abrikosov1988}. For junctions with 
identical superconducting electrodes those normalization factors cancel each other out, and the boundary condition for
the standard GL order parameter $\Psi$ coincides with \eqref{Jos15}
\be
\dfrac{\sigma}{G_N}\left(\dfrac{d\Psi}{dX}\right)_\pm=\Psi_+-\Psi_- .
\label{Jos13}
\ee

One can also introduce in \eqref{Jos8} and \eqref{Jos13} the dimensionless coordinate $x=X/\xi$, where $\xi$ is the 
superconductor coherence length, and rewrite the equalities as
\be
\left(\dfrac{d\Psi}{dx}\right)_\pm=\frac{\xi g_J}{K}\left(\Psi_+-\Psi_-\right)=g_\ell\left(\Psi_+-\Psi_-\right),
\label{Jos8b}
\ee
\be
\left(\dfrac{d\Psi}{dx}\right)_\pm=\dfrac{\xi G_N}{\sigma}\bigl(\Psi_+-\Psi_-\bigr) \,.
\label{Jos18}
\ee

Comparing the microscopic and the GL boundary conditions \eqref{Jos13} and \eqref{Jos8}, as well as the associated
equalities \eqref{Jos8b} and \eqref{Jos18}, leads to the following expressions 
\be
g_J=\dfrac{KG_N}{\sigma}, \qquad g_\ell=\dfrac{\xi G_N}{\sigma}.
\label{Jos17}
\ee

Therefore, the condition $\xi G_N/\sigma\ll1$, which is known to allow the junction description in the tunneling limit, 
is rewritten in terms of $g_\ell$ as $g_\ell\ll1$.

Microscopic expressions for $g_J$, $g_\ell$ can also be obtained by comparing microscopic results for the Josephson 
current near $T_c$ with the corresponding formulas of the GL theory. In the absence of surface pair breaking, the GL 
expression for the critical current in the tunneling limit $j_c=\frac{4|e||a|}{\hbar b}g_J$, that applies to symmetric 
SIS tunnel junctions, follows from \eqref{Jos2} after substituting there the conventional bulk GL expression 
$\left|\Psi_\pm\right|^2=|a|/b$ for the equilibrium order parameter via the standard GL coefficients (see, e.g., 
\eqref{Fb1}). On the other hand, the microscopic Ambegaokar-Baratoff formula near $T_c$ for the same system can be
represented as $j_c=\pi|\Delta|^2G_N/4|e|T_c$, where the BCS gap function in the bulk near $T_c$ is 
$|\Delta|^2=8\pi^2T_c(T_c-T)/(7\zeta(3))$~\cite{Ambegaokar1963,*Ambegaokar1963err}. Thus, equating the two expressions
and using the relation $|a|=\alpha |T_c-T|/T_c$ one gets
\be
g_J=\dfrac{\pi^3\hbar bT_c G_N}{14\zeta(3)e^2\alpha}.
\label{Jos16}
\ee 

Eq. \eqref{Jos16} can be specified further with the Gor'kov's microscopic formulas for $b/\alpha$
in the clean and dirty limits~\cite{Gor'kov1959ar,Gor'kov1959br}. Substituting the simplest expression 
$G_N=e^2k_F^2\overline{\cal D}/4\pi^2\hbar$, that connects the junction conductance with the averaged transmission 
coefficient $\overline{\cal D}$, one gets for pure junctions $g_\ell=3\pi^2\overline{\cal D}\xi(T)/(14\zeta(3)\xi_{0})=
1.76\overline{\cal D}\xi(T)/\xi_{0}$. Here $\xi_0=\hbar v_F/\pi T_c$ is the zero-temperature coherence length and $v_F$,
$k_F$ are the Fermi velocity and wave vector. Based on \eqref{Jos16} and the microscopic expression for $b/\alpha$ for 
dirty junctions and using the relations $K=\hbar^2/4m$, $\sigma=n_ee^2l_{\text{imp}}/p_f$, where $l_{\text{imp}}$ is the
mean free path, one obtains the same results \eqref{Jos17} derived above with the Kupriyanov-Lukichev boundary
conditions. Further substitution of the expression for $G_N$ in \eqref{Jos17} results in 
$g_\ell=0.75\overline{\cal D}\xi(T)/\ell$. 

The estimates made above also agree with the microscopic formulas obtained long ago for boundary conditions 
for the order parameter near $T_c$ in the tunneling limit~\cite{Galaiko1969,Bratus1977,Svidzinskii1982}. The 
estimate $g_\ell\sim \overline{\cal D}\xi(T)(l_{\text{imp}}^{-1}+ \xi_0^{-1})$, that sometimes is identified as the 
effective transparency~\cite{Bezuglyi1999,Bezuglyi2005}, applies not only to junctions in the tunneling limit 
$g_\ell\ll 1$ but also to those with moderate transparency not too close to unity. As the ratio $\xi(T)/l$ in dirty
superconductors can be $\sim 100$ even at low temperatures, the quantity $g_\ell\sim\overline{\cal D}\xi(T)/l$ can vary,
for small and moderate transparencies, from vanishingly small values in the tunneling limit to those 
well exceeding $100$ near $T_c$, when a pronounced anharmonic behavior of the Josephson current is known to take
place~\cite{Kupriyanov1992,Barash2012,Barash2012_3,Barash2014_3}. In highly transparent junctions, for which 
$1 - {\cal D}\ll1$, the parameter $g_\ell\propto (1 - {\cal D})^{-1}$ can be very large, as follows from microscopic 
results for the Josephson current in planar junctions with thin 
interlayers~\cite{Galaiko1969,Bratus1977,Ivanov1981,Svidzinskii1982,Kupriyanov1992} (see 
also~\cite{Golubov2004,Geshkenbein1988}).

\section{Internal Phase Differences in \mbox{SINIS} Double Junctions}
\label{sec: sinis}

\subsection{Basic Equations}
\label{subsec: model1}

This section presents a theory of the reduced range of changes and unconventional behaviors of internal phase 
differences $\chi_{1,2}$ across the two interfaces in symmetric \mbox{SINIS} double tunnel junctions, that occur under 
the effect of controlled variations of either the external phase difference $\phi$ between superconducting terminals, or the
quantities $\chi_{1,2}$ themselves.

\begin{figure}[t]
\centering
\includegraphics*[width=.4\textwidth,clip=true]{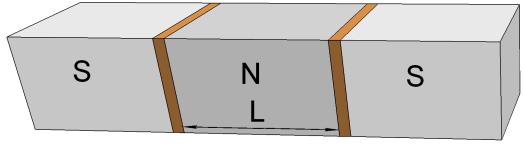}
\caption{Schematic diagram of the \mbox{SINIS} junction. Adopted from Ref.~\cite{Barash2019}.} \label{fig1}
\end{figure}

The interfaces are set upon the end faces of the central normal metal lead of length $L$ (see Fig.\,\ref{fig1}). A 
one-dimensional spatial dependence of the order parameter is considered, as it is assumed that the transverse dimensions 
of the electrodes are significantly smaller than the superconductor coherence length $\xi$ and the decay length $\xi_n$ 
in the normal metal electrode. The system's free energy consists of the bulk and interface terms 
${\cal F}=\sum{\cal F}_{p}+{\cal F}_{n}+{\cal F}^{\text{int}}_{\frac{L}2}+{\cal F}^{\text{int}}_{-\frac{L}2}$. Here 
$p=1,2$ correspond to the external superconducting electrodes, while subscript $n$ - to the central normal metal lead.

For the central electrode, one can write the following bulk GL free energy per unit area of the cross section
\be
\!\!{\cal F}_n\!=\!\!\!\int\nolimits_{-L/2}^{L/2}\!\!\!dX\!
\left[\!K_n\left|\dfrac{d}{dX}\Psi(X)\right|^2\!\!\!+a_n\left|\Psi(X)\right|^2\!\!+\dfrac{b_n}{2}
\left|\Psi(X)\right|^4\right]\!,
\label{Fb1b}
\ee
where $K_n,\,a_n,\,b_n>0$ and the interfaces are situated at $X=\pm L/2$. The expressions for ${\cal F}_{1,2}$ 
can be obtained from \eqref{Fb1b} by substituting $K,\,-|a|,\,b$ for $K_n,\,a_n,\,b_n$, respectively, and replacing the 
integration period $(-L/2,L/2)$ by $(-\infty,-L/2)$ or $(L/2,\infty)$ for $p=1$ or $2$.

The interface terms in the free energy per unit area at $X=\pm L/2$ are similar to \eqref{Jos12}:
\be
{\cal F}^{\text{int}}_{\pm\frac{L}2}=g_{J}\Bigl|\Psi_{\pm\left(\frac{L}{2}+0\right)}-
\Psi_{\pm\left(\frac{L}{2}-0\right)}\Bigr|^2 
+g\Bigl|\Psi_{\pm\left(\frac{L}{2}+0\right)}\Bigr|^2
+g_n\Bigl|\Psi_{\pm\left(\frac{L}{2}-0\right)}\Bigr|^2.
\label{fint1b}
\ee

The GL equation for the normalized absolute order-parameter value can be written as 
\be\left\{
\begin{aligned}
&\dfrac{d^2f}{dx^2}-\dfrac{i^2}{f^3}-f-f^3=0,\quad |x|<l/2,\\  \\
&\dfrac{d^2f}{dx^2}-\dfrac{K_n^2}{K^2}\cdot\dfrac{i^2}{f^3}+\dfrac{|a|K_n}{a_nK}f-
\dfrac{b K_n}{b_nK}f^3=0,\quad |x|>l/2.
\end{aligned}
\right.
\label{rlambda11}
\ee
Here the dimensionless quantities $f$ and $x$ are defined as $\Psi\!=\!(a_n/b_n)^{1/2}fe^{\mathtt{i}\alpha}$,
$x\!=\!X/\xi_n(T)$. Also $l\!=\!L/\xi_n$, $\xi_n(T)\!\!=\!(K_n/a_n)^{1/2}$ and the dimensionless 
current density is $i\!=\!\frac{2}{3\sqrt{3}}(j\big/j_{\text{dp}})$, where $j_{\text{dp}}\!=\!
\bigl(8|e|a_n^{3/2}K_n^{1/2}\bigr)\big/\bigl(3\sqrt{3}\hbar b_n\bigr)$. One also assumes $a_n\sim|a|$ that
makes possible a joint description of the normal metal and superconducting leads within the GL 
approach~\cite{Abrikosov1988}.

The boundary conditions for the complex order parameter, obtained from \eqref{Fb1b}, \eqref{fint1b} on the 
interfaces at $x=\pm l/2$, are reduced to following equalities for real quantities
\be\left\{
\begin{aligned}
&\left(\dfrac{df}{dx}\right)_{\pm(l/2-0)}=\mp\bigl(g_{n,\delta}+g_{\ell}\bigr)f_-\pm g_{\ell}f_+\cos\chi,
\\
&\dfrac{K}{K_n}\left(\dfrac{df}{dx}\right)_{\pm(l/2+0)}=\pm(g_\delta+g_{\ell})f_+\mp g_{\ell}f_-\cos\chi,
\end{aligned}
\right.
\label{rlambda30a}
\ee
\be
i=-\,f^2\dfrac{d\alpha}{dx}=g_{\ell}f_{-}f_{+}\sin\chi.
\label{jcurbc}
\ee
Here the subscripts $+$ and $-$ in $f_\pm$ correspond to the superconducting and normal metal banks respectively. The dimensionless
coupling constants are
\be
g_\ell=g_J\xi_n(T)/K_n, \quad g_{n,\delta}=g_n\xi_n(T)/K_n, \quad g_\delta=g\xi_n/K_n
\label{couplconst}
\ee
and the symmetric solutions $f(x)=f(-x)$ are considered.

\subsection{Phase Relations in \mbox{SINIS} Junctions} 
\label{subsec: phaserel}

If a superconducting order parameter $f_+$ is present on one of the sides of a thin interface and the condition
$g_\ell\cos\chi>0$ is satisfied, then the bilinear part of the Josephson free energy $\propto-2g_\ell f_-f_+\cos\chi$ 
(see \eqref{fint1b} and \eqref{Jos1}) decreases with an appearance of a small nonzero order parameter $f_-$ on the other
side. This results in the proximity effect of the Josephson origin related to the term with $\cos\chi$ in the boundary 
conditions \eqref{rlambda30a}. Therefore, for superconductivity to emerge on the opposite interface side in the normal 
metal lead, in the presence of $0$-junctions considered below ($g_\ell>0$), the internal phase difference $\chi$ must 
take its values within the proximity-reduced range $|\chi|\le\chi_{\text{max}}(l)<\frac{\pi}{2}$ defined modulo $2\pi$. 
Outside the range, the Josephson coupling would prevent superconductivity to appear in the normal metal lead.

\begin{figure*}%
\centering
\begin{subfigure}{0.4\linewidth}
\includegraphics[height=0.6\textwidth]{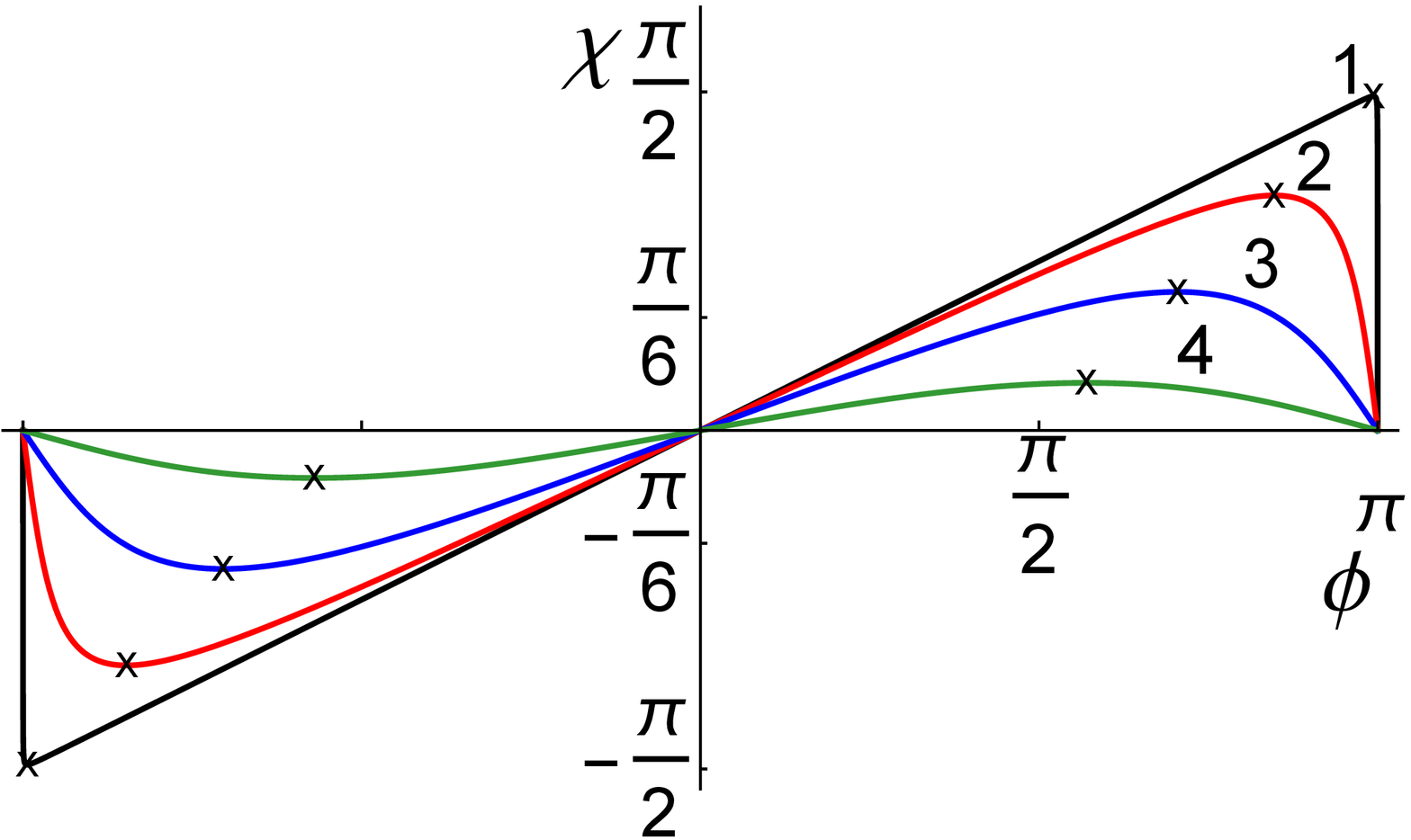}
\caption{}
\label{fig2a}
\end{subfigure}
\hspace{25mm}
\begin{subfigure}{0.4\linewidth}
\includegraphics[height=0.6\textwidth]{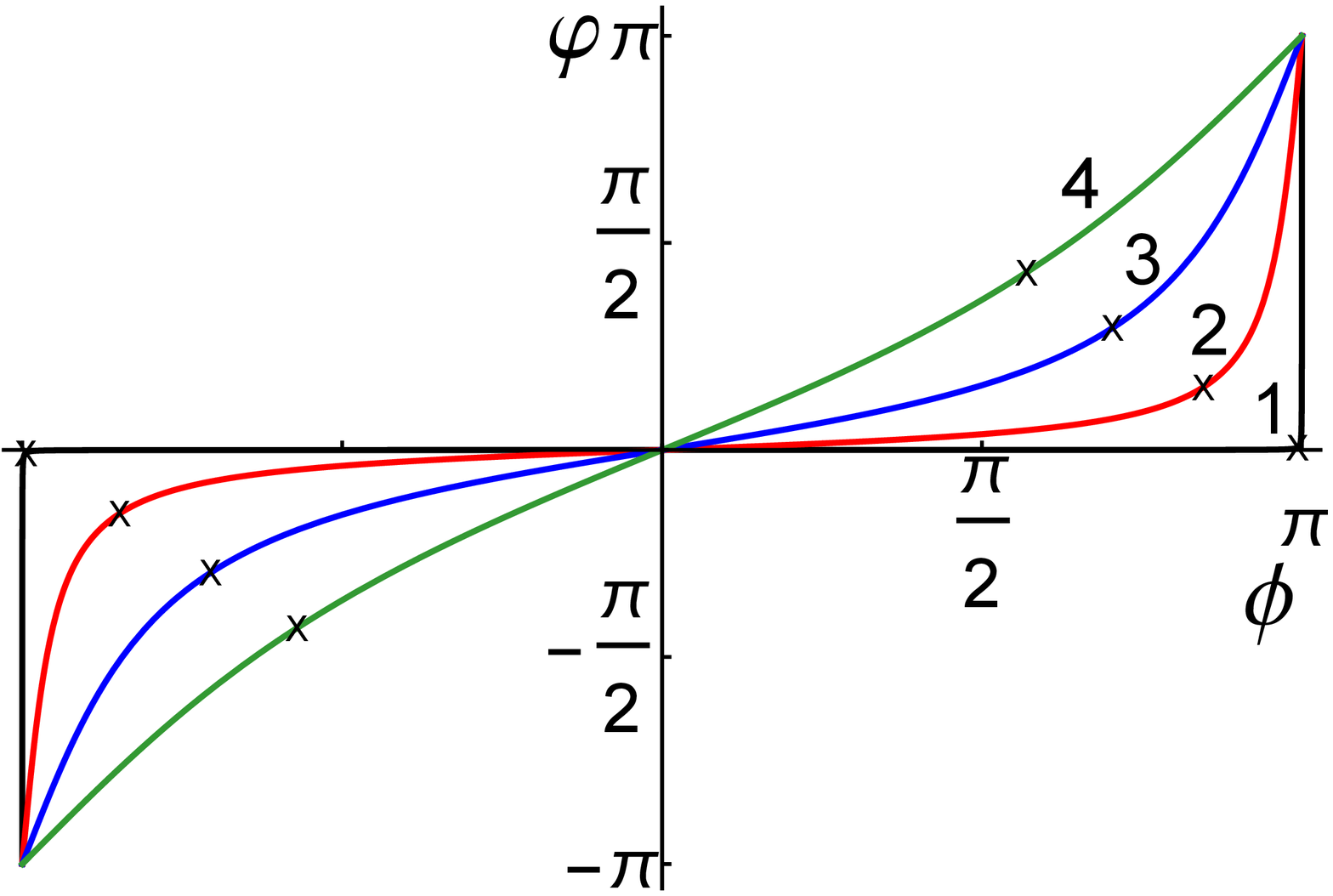}
\caption{}
\label{fig2b}
\end{subfigure}
\caption{The internal phase difference $\chi$ (a) and the phase incursion $\varphi$ (b)
as functions of the external phase difference $\phi$ taken at various central lead's lengths $l$:\,
(1)\, $l=0.02$\,\, (2)\, $l=0.5$\,\, (3)\, $l=1.1$\,\, (4)\, $l=2.2$. Adopted from Ref.~\cite{Barash2019}.}%
\label{fig:chi-varphi-phi-l}%
\end{figure*}

The internal phase difference $\chi=\chi_{1,2}$ and the phase incursion $\varphi$ taken at various $l$, are shown 
in \cref{fig2a,fig2b}, respectively, as functions of the external phase difference $\phi$.
The numerical results shown have been obtained by evaluating the GL equations' solutions with 
$g_\ell\!=\!g_\delta\!=\!0.01$, $g_{n,\delta}\!=\!0$, $K=K_n$, $|a|=a_n$ and $b=b_n$. \!The analytical solutions, 
presented in Subsec.~\ref{subsec: tunnel} for tunnel \mbox{SINIS} junctions, approximate almost perfectly the 
quantities $\chi(\phi)$ and $\varphi(\phi)$ for the parameter set considered, with deviations from the numerical results 
that are indiscernible in \cref{fig2a,fig2b}.

A simple relationship $\chi(\phi)=\frac{\phi}2$, which follows from the equality $\phi=2\chi+\varphi$ in the case of 
negligibly small phase incursion, results in the variation range $|\chi|\le\frac{\pi}{2}$ for $|\phi|\le\pi$. As seen in
the curves 1 in \cref{fig2a,fig2b}, such behavior occurs at sufficiently small lengths $l\ll1$, except 
for small vicinities of $\phi=\pm\pi$. The curves 2-4 show that the internal phase difference $\chi$ substantially 
contributes to the phase relations in a wide range of $\phi$ and at mesoscopic lengths $l\alt1$, while $\varphi$ is of
importance at $l\agt1$.

Due to a spatially constant supercurrent occurring in the quasi-one-dimensional problem considered, a
local decrease of the condensate density leads to the increase of the superfluid velocity, i.e., of 
the gradient of the order parameter phase. In this regard, a spatial decay of the proximity-induced 
Cooper pair density that takes place inside the central electrode as a distance from the nearest 
interface increases, plays an important role. This is the origin of a noticeable phase incursion 
taking place in a wide range of the phase difference $\phi$ in \mbox{SINIS} double junctions with a 
central lead of a mesoscopic length $l$. Smaller local values of $f$ near the lead's center at larger
$l$ increases the phase incursion $\varphi$ at a given $\phi$, while the range of variation
of $\chi$ diminishes $|\chi|\le\chi_{\text{max}}(l)$. In accordance with the result \eqref{chimax} 
of Subsec.~\ref{subsec: tunnel}, $\chi_{\text{max}}(l)\approx\arccos(\tanh l)$. When $l\gg1$, the 
order parameter is especially small in the depth of the central electrode and the phase incursion 
$\varphi$ dominates $\chi$ in the equality $\phi=2\chi+\varphi$, while $|\chi|$ is greatly reduced.

As seen in \cref{fig2a}, the internal phase difference $\chi$ is a nonmonotonic function of the external phase 
difference $\phi$. With $\phi$ changing over the period, $\chi$ passes through its proximity-reduced region twice, there
and back. Two different values of $\phi$, that correspond to one and the same $\chi$, and to different phase incursions,
belong to the two solutions of the GL equation for the absolute value of the order parameter, taken at a given
$\chi$. In all the figures, the dots marked with crosses indicate the points of contact of the two solutions, i.e., the 
quantities taken at $\chi=\pm\chi_{\text{max}}(l)$.

\subsection{The Proximity-Induced Phase Dependent Order Parameter}
\label{subsec: orderparchi}

The nonlinear term $i^2f^{-3}\propto v_s^2(x)f(x)$, that takes into account an influence of the conserved supercurrent
on the proximity effect, i.e., on the induced order parameter in the normal metal electrode, cannot, as a rule, be
neglected in \eqref{rlambda11} in comparison with the linear term. When $\phi$ is close to $\pi$, it dominates the
latter in the depth of the central lead (see \eqref{pap21} and its discussion in Subsec.~\ref{subsec: tunnel}). Thus,
the GL equation \eqref{rlambda11} is a nonlinear one even if the conventional cubic term is negligible in the problem in
question. Although in the latter case the equation for the complex order parameter amplitude is still linear, the
nonlinearity appears in describing the order parameter's absolute value and phase. As a result of the nonlinearity, two
basic solutions for $f$ at a given $\chi$ will be shown to take place in the problem even for very small order parameter
values. However, in the absence of a sizable spatially dependent gauge invariant gradient of the order parameter phase
that is linked with the order parameter absolute value by the current conservation condition, the linearization
represents the simplest and most effective way of solving the problems of $H_{c2}$ and
$H_{c3}$~\cite{Abrikosov1957,DeGennes1963,Tinkham1996} as well as of the proximity effects near
superconductor-normal metal boundaries~\cite{DeGennes1969,Abrikosov1988}.

\begin{figure*}%
\centering
\begin{subfigure}{.4\linewidth}
\includegraphics[height=0.6\textwidth]{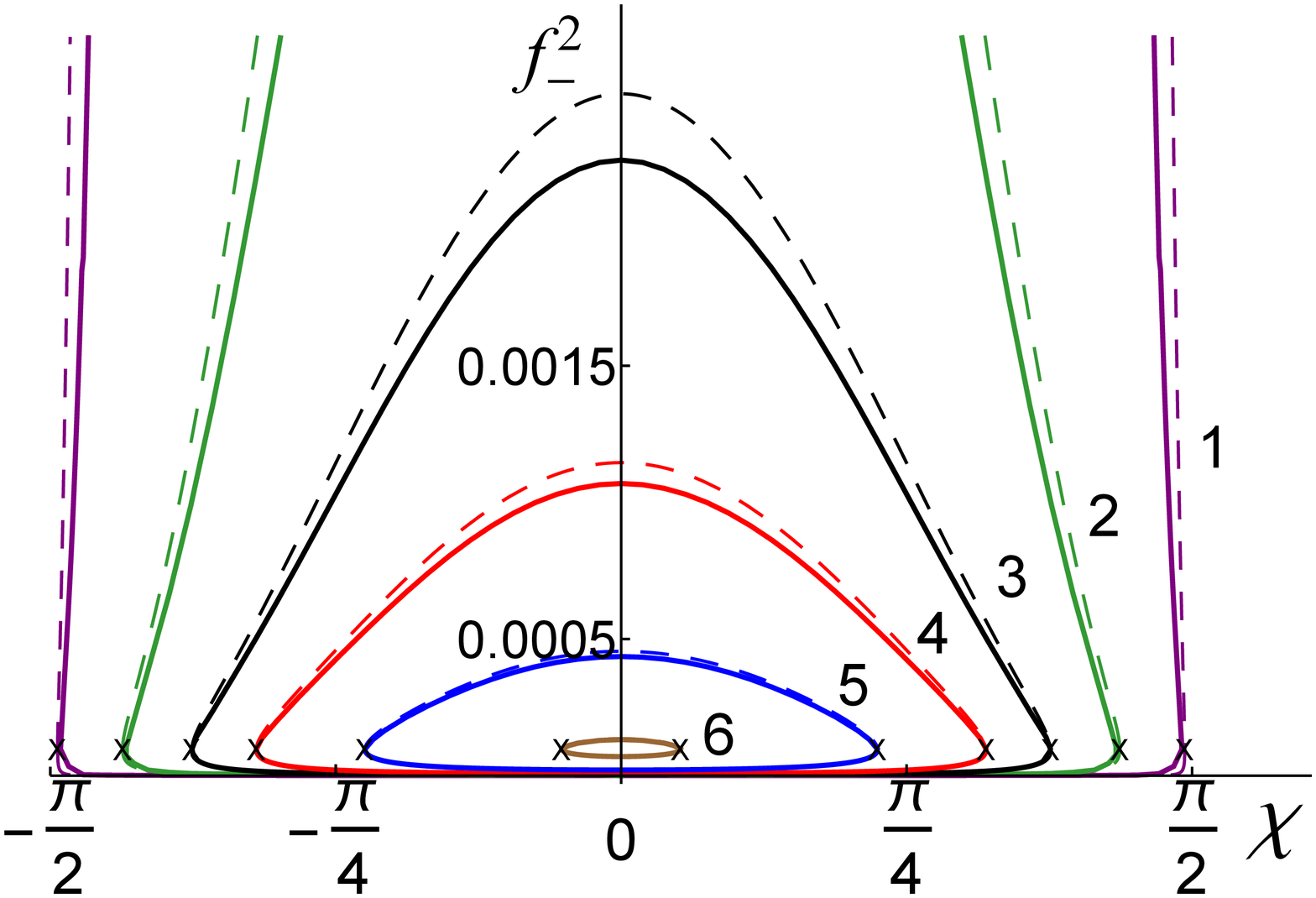}
\caption{}
\label{fig3a}
\end{subfigure}
\hspace{25mm}
\begin{subfigure}{.4\linewidth}
\includegraphics[height=0.6\textwidth]{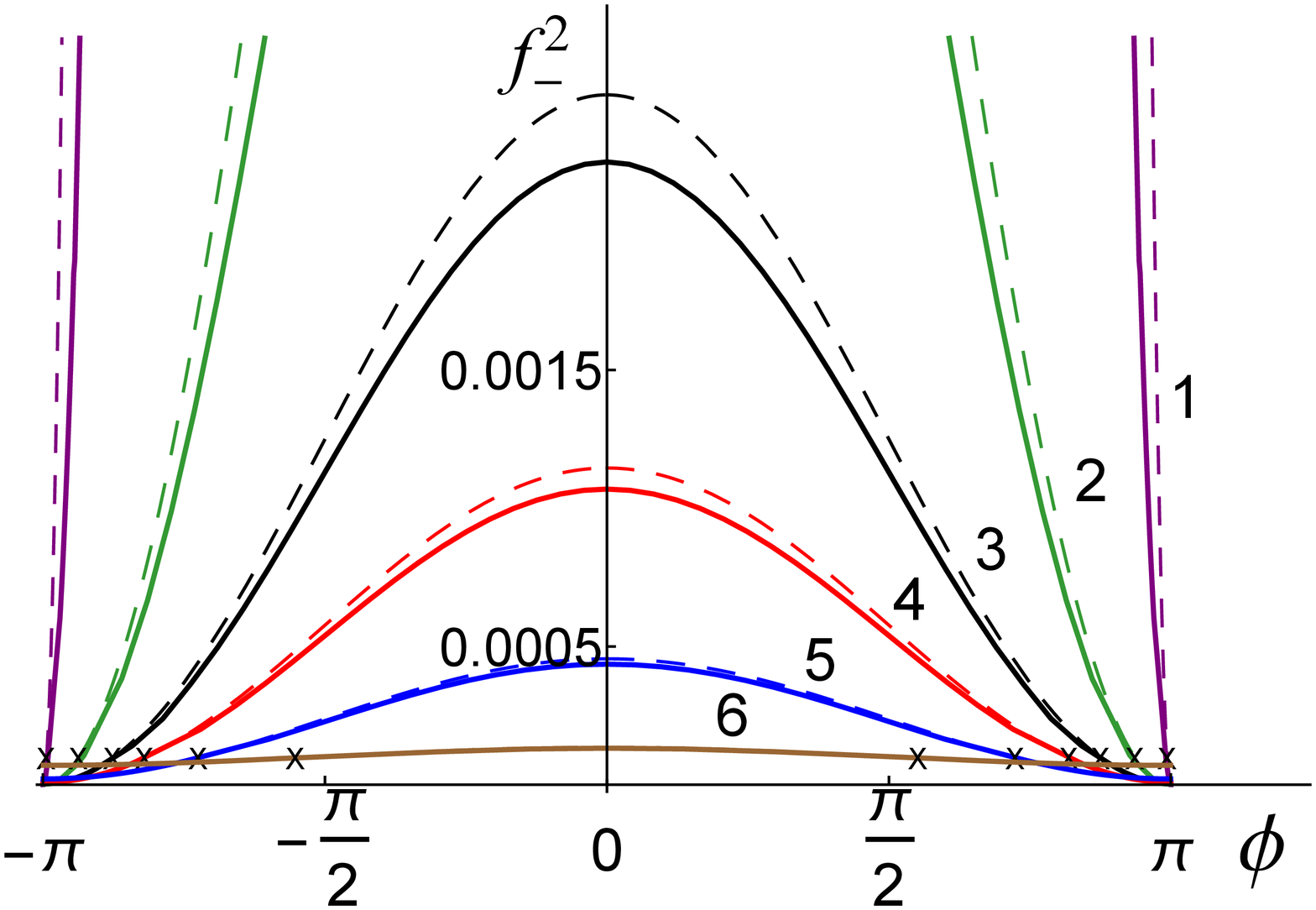}
\caption{}
\label{fig3b}
\end{subfigure}
\caption{The quantity $f_{-}^2$ as (a) a double-valued function of $\chi$  and (b) a single-valued 
function of $\phi$ taken at various $l$:\, 
(1)\, $l=0.02$\,\, (2)\, $l=0.2$\,\, (3)\, $l=0.4$\,\, (4)\, $l=0.6$\,\, (5)\, $l=1$\,\, 
(6)\, $l=2.5$. Adopted from Ref.~\cite{Barash2019}.}%
\label{fig:t0-chi-phi-l}%
\end{figure*}

\Cref{fig3a,fig3b} show the normalized order parameter absolute value squared $f^2_-$, taken on the end
face of the central electrode at various $l$, as a function of $\chi$ and $\phi$, respectively. Solid curves 
give the numerical results obtained within the same framework as in Fig.~\ref{fig:chi-varphi-phi-l}. Dashed curves 
depict the analytical results of Subsec.~\ref{subsec: tunnel}, that assume the conditions 
$f_-\!\sim \!g_\ell f_+\!\ll\!1$, $g_{n,\delta}\alt g_\ell$ and lead to \eqref{pap12}, 
\eqref{lsmall132} - \eqref{lsmall134}, \eqref{lsmall168}. They approximate the numerical results reasonably well. Unlike
the phase relations in Fig.~\ref{fig:chi-varphi-phi-l}, the solid and dashed curves in Fig.~\ref{fig:t0-chi-phi-l} can 
be, for the most part, clearly distinguished. The double-valued behavior shown in \cref{fig3a} is described by the
two solutions for $f^2_-$ adjoining at $\chi\!=\!\pm \chi_{\text{max}}(l)$. The first solution has the maximum value and
the second one the minimum at $\chi\!=\!0$ at a fixed $l$. A similar behavior takes place at $l\to0$ at a fixed $\chi$, 
where the minimum is zero.

If $\chi$ were the control parameter in an experiment, the first solution would represent the stable states and the 
second one the metastable states. However, the external phase difference $\phi$ is usually fixed experimentally. After 
switching over from $\chi$ to $\phi$, the two solutions occupy different regions within the period, adjoining at the 
points $\phi=\pm\phi_*(l)$. They jointly form the continuous single-valued order parameter behavior $f_-(\phi,l)$ shown 
in \cref{fig3b}. The first solution incorporates $|\phi|\in(0,\phi_*(l))$ while the second one is in 
$|\phi|\in(\phi_*(l),\pi)$. Here $\phi_*(l)\approx\frac{\pi}{2}+\arcsin(\frac{1}{\cosh l})$, in accordance with 
\eqref{pap30}, and the adjoining regions do not overlap due to a pronounced phase incursion taking place at small $f$. 
The curves' crossing at small $f_-$, shown in \cref{fig3b}, is a consequence of different behavior of the two 
solutions with increasing $l$. Phase-slip processes in the central lead can take place at $\phi=\pm\pi$ and at arbitrary 
$l$, as in this case $f$ vanishes at $x=0$~\cite{Fink1976,Giazotto2017} (see also \eqref{pap21}). 
This results in a noticeable phase incursion in immediate vicinities of $\phi=\pm\pi$ even at small 
values of $l$.

The quantity $f_+$ changes weakly with $\chi$ and $l$: $f_+^2\in(0.972,0.978)$ for the whole parameter set used in the 
figures. For tunnel interfaces the estimate $f_-\sim g_\ell f_+\ll1$ holds at $g_{n,\delta}\alt g_\ell$, with the 
exception of the first solution at sufficiently small $l$. In the limit of small $l$, the first solution satisfies the 
relation
\be
f_-\!=\!\frac{g_\ell\cos\chi}{g_\ell+g_{n,\delta}}f_+,
\label{lto0}
\ee
which assumes $\cos\chi\ge0$ and approximately describes the dependence of $f_-$ on $\chi$ and also applies to 
\mbox{SISIS} junctions~\cite{Barash2018}. 

If $g_{n,\delta}\alt g_\ell$ and $\cos\chi\sim1$, it follows from \eqref{lto0} $f_-\sim f_+$. In the opposite case
$g_{n,\delta}\gg g_\ell$ one gets $f_-\ll f_+$. Since $g_\ell$ for tunnel interfaces is proportional to the 
transmission coefficient $g_\ell\propto{\cal D}$ (see Subsec.~\ref{subsec: micro}), one obtains from \eqref{lto0} 
$f_-\propto{\cal D}$, if $g_\ell\ll g_{n,\delta}$, and the ${\cal D}$-independent quantity $f_-$ for 
$g_\ell\gg g_{n,\delta}$. The second solution vanishes in the limit $l\to0$, and satisfies the relation 
$f_-=g_\ell f_+\tanh\frac{l}{2}$ at arbitrary $l$ and $\chi=0$. The two solutions for large $l$ coincide and 
the relation at $\chi=0$ is $f_-=g_\ell f_+$ (see Supplemental Material in Ref.~\cite{Barash2019}). 

\begin{figure*}%
\centering
\begin{subfigure}{.4\linewidth}
\includegraphics[height=0.6\textwidth]{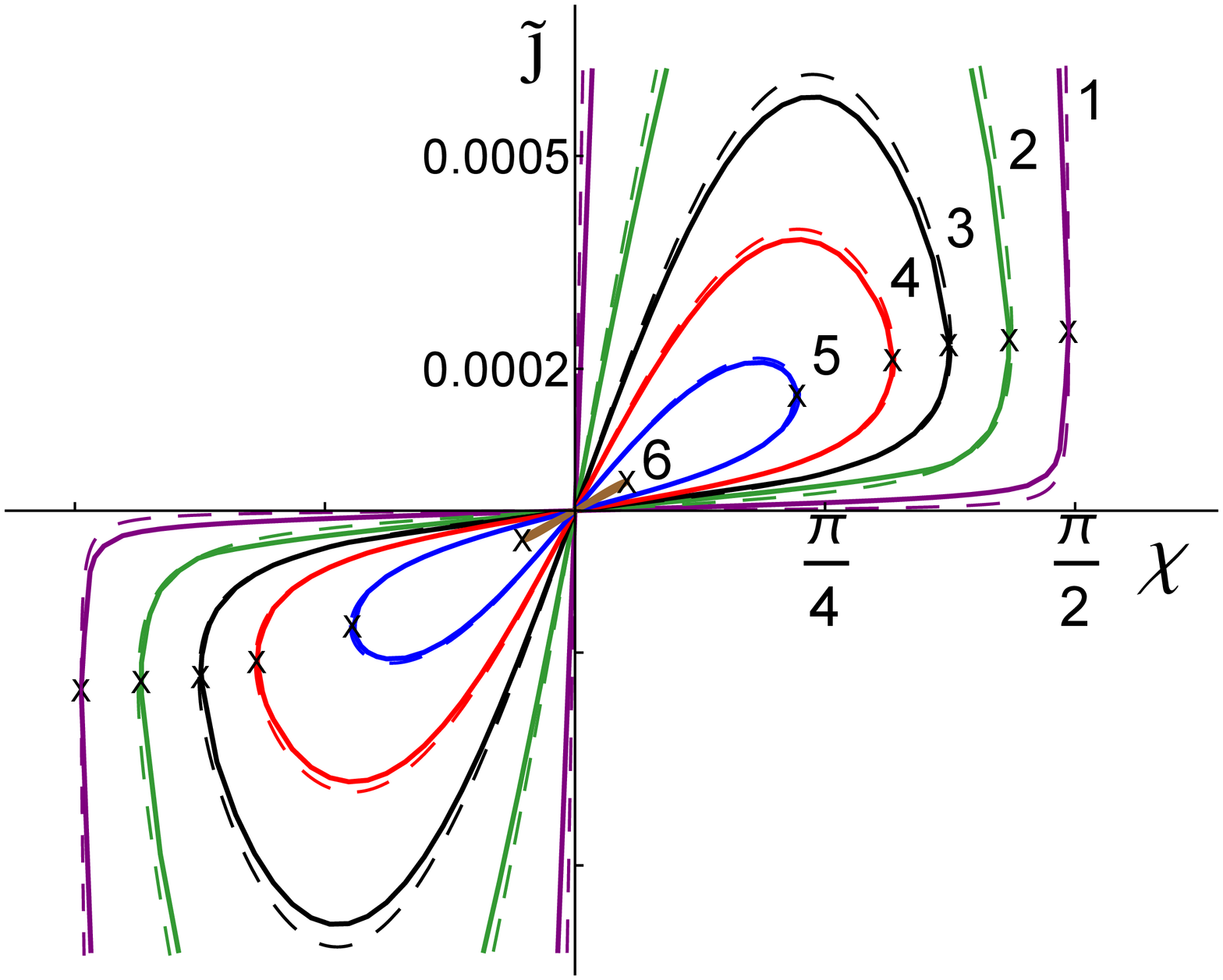}
\caption{}
\label{fig4a}
\end{subfigure}
\hspace{20mm}
\begin{subfigure}{.4\linewidth}
\includegraphics[height=0.6\textwidth]{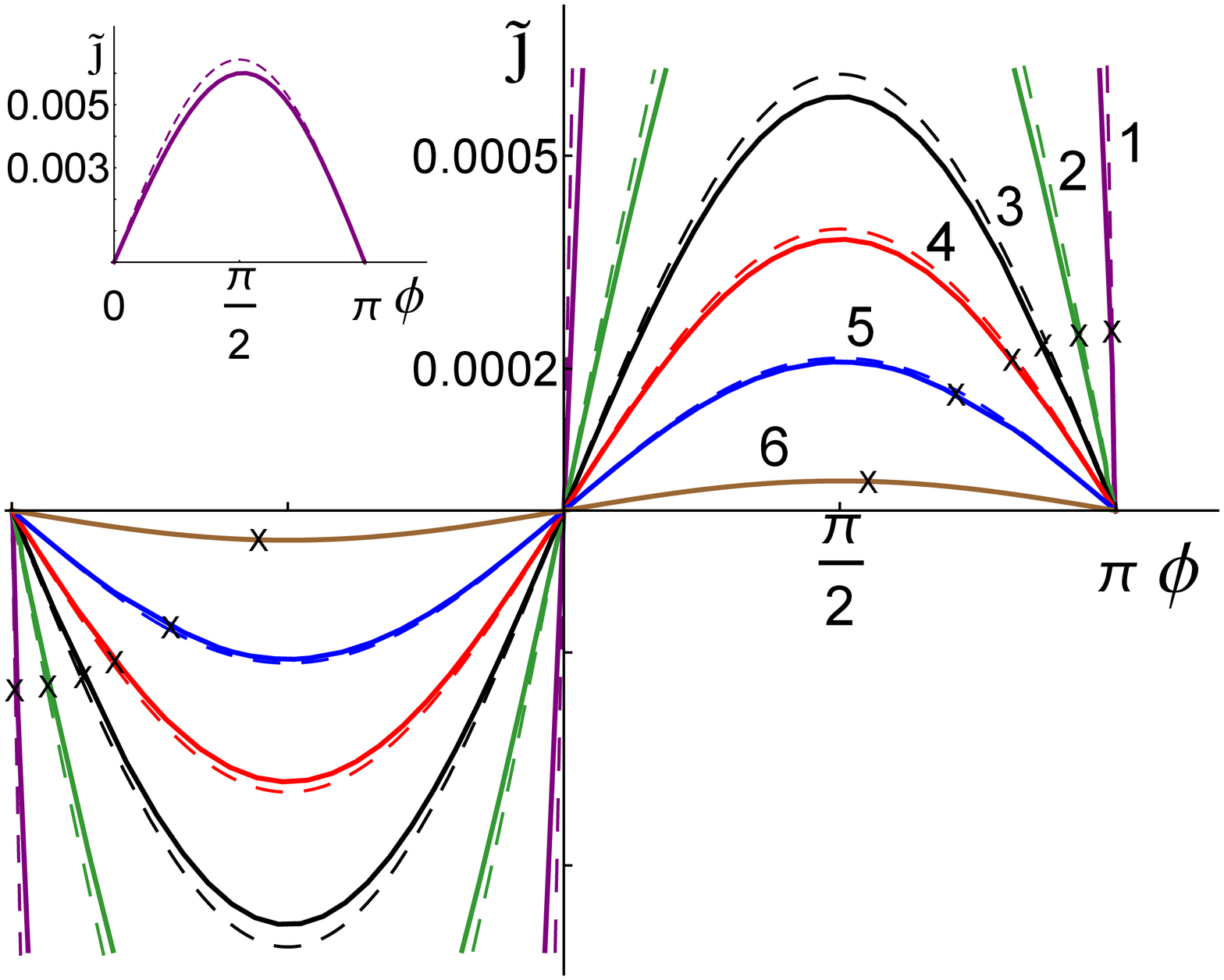}
\caption{}
\label{fig4b}
\end{subfigure}
\caption{Normalized supercurrent as (a) a double-valued function of $\chi$ and (b) a single-valued 
function of $\phi$ taken at various $l$: \, (1)\, $l=0.02$\,\, (2)\, $l=0.2$\,\, 
(3)\, $l=0.4$\,\, (4)\, $l=0.6$\,\, (5)\, $l=1$\,\, (6)\, $l=2.5$.\enspace
Inset: The supercurrent at l=0.02 (solid line) and its analytical description at small $l$ (dashed line). Adopted 
from Ref.~\cite{Barash2019}.}%
\label{fig:cur-chi-phi-l}%
\end{figure*}

\subsection{Dependence of the Supercurrent on Internal Phase Differences in \mbox{SINIS}}
\label{subsec: curchi}

The normalized supercurrent $\tilde{\jmath}=j\big/j_{\text{dp}}$ as a function of $\chi$ and $\phi$ is depicted at
various $l$ in \cref{fig4a,fig4b}, respectively. With varying $\chi$, the curves have the form of a 
double loop looking like a sloping figure eight composed of the two solutions. By contrast, after switching from $\chi$ 
to $\phi$ the transformed curves correspond to the conventional current-phase relation. The dashed curves are 
based on the approximate analytical results of Subsec.~\ref{subsec: tunnel}. They deviate by several percent from
the numerical solutions (solid curves) having the sinusoidal shape in \cref{fig4b}.

A role of the phase incursion in producing the obtained phase-dependent behavior can be explained as follows. 
The factor $f_-$ in the supercurrent $i=g_{\ell}f_{-}f_{+}\sin\chi$ is most sensitive to the proximity effect.
If $\varphi$ were negligibly small, the value of the external phase difference $\phi\!=\!\pi$ would correspond to 
$\chi\!=\!\frac{\pi}2$. Since the bilinear part of the Josephson free energy $\propto-2g_\ell f_-f_+\cos\chi$ and the 
corresponding proximity effect vanish at $\chi\to\frac{\pi}{2}$, one obtains $f_-\!\!\to0$ that entails the zeroth 
supercurrent at $\phi=\pi$. At the same time, small values of $f_-$ in the vicinity of $\phi=\pi$ result in a noticeable
phase incursion. In view of the relation $\chi=(\phi-\varphi)/2$, the phase incursion diminishes the variation 
range $|\chi|\le\chi_{\text{max}}(l)<\frac{\pi}{2}$ that prohibits $\chi$ to reach $\frac{\pi}{2}$ at any nonzero $l$. 
Instead, two solutions of the GL equation provide a return passage for $\chi$, from $0$ to 
$\chi_{\text{max}}(l)$ and back, when $\phi$ changes over $(0,\pi)$. This results in the correspondence of $\chi=0$ to 
both $\phi=0$ and $\phi=\pi$.  

Since the analytical results of Subsec.~\ref{subsec: tunnel} can only be justified for sufficiently small $g_\ell$, 
the small Josephson coupling constant $g_\ell=0.01$ has been chosen to demonstrate a quantitative agreement with the 
numerical data. The chosen $g_\ell$ leads to small values of the order parameter $f_{-}$ and the supercurrent, which 
pertain to the second solution and are depicted in Figs.\,\ref{fig:t0-chi-phi-l} and \ref{fig:cur-chi-phi-l} below the 
points marked with crosses. The effects in question increase with $g_\ell$ and remain qualitatively the same for 
$g_\ell\alt 1$. For example, for $g_\ell=0.1$ instead of $g_\ell=0.01$, the typical values of $f_-^2$ and 
$\tilde{\jmath}$ increase in about $50-100$ times.

The first solution is substantially modified at small length of the central lead $l\alt 2g_\ell (b_n|a|/ba_n)^{1/2} \ll1$. 
The corresponding analytical description should be based on the relation \eqref{lto0} rather than on 
$f_-\sim g_\ell f_+\ll1$ (see Supplemental Material in Ref.~\cite{Barash2019}). The solid curve 1 as a whole ($l=0.02$) 
is shown in the inset in \cref{fig4b}. It is well approximated by the analytical results. Under the condition 
$g_{n,\delta}\ll g_\ell$, one obtains $j\propto {\cal D}$ from \eqref{lto0}. However, the supercurrent dependence on the
transparency is gradually transformed into ${\cal D}^2$ with increasing $\phi$ up to about $\phi_*$ at a fixed small $l$,
when the relation $f_-\sim g_\ell f_+\ll1$ comes into play along with the increase of the phase incursion. The second
solution always results in $j\propto{\cal D}^2$. The crossover obtained is a fingerprint of the phase-dependent 
proximity effect of the Josephson origin that results in an unconventional behavior of internal phase differences in 
\mbox{SINIS} junctions. A similar behavior of a supercurrent also occurs with the increasing distance $l$ at a fixed 
$\phi$~\cite{Kupriyanov1988,Kupriyanov1999,Golubov2000,Golubov2004}.

\subsection{Analytical Results for Tunnel \mbox{SINIS} Junctions}
\label{subsec: tunnel}

This section focuses on the analytical results used above for plotting the dashed curves in 
Figs.~\ref{fig:chi-varphi-phi-l} - \ref{fig:cur-chi-phi-l}. For a more complete analytical consideration of the problem
see Supplemental Material in Ref.~\cite{Barash2019}.

The GL equation and boundary conditions, applied to the complex order parameter with one dimensional spatial dependence,
lead not only to the equations \eqref{rlambda11}, \eqref{rlambda30a}, but also to additional equalities, that ensure the
conservation of the current through both the superconducting leads and the interfaces. The equalities can be written in 
the form of \eqref{jcurbc} and as
\be
i^2=\dfrac{Kb}{K_nb_n}\left(\dfrac{|a|b_n}{a_nb}-f_\infty^2\right)f_\infty^4.
\label{ascur1}
\ee
The first expression in the right-hand side of \eqref{jcurbc} is the normalized standard GL expression for the 
supercurrent, which is valid everywhere inside the leads. The second expression in \eqref{jcurbc} is the supercurrent 
value at the boundary, as it follows from the boundary condition for the complex order parameter. The expression 
\eqref{ascur1} for $i^2$ is obtained from \eqref{rlambda11} in the limit $x\to\infty$, in view of the relation 
$df/dx\to0$.

The asymptotic order parameter value generally depends on the supercurrent.
Equating \eqref{jcurbc} and \eqref{ascur1} results in
\be
\dfrac{Kb}{K_nb_n}\left(\dfrac{|a|b_n}{a_nb}-f_\infty^2\right)f_\infty^4=
g_\ell^2f_-^2f_+^2\sin^2\chi.
\label{sol3b}
\ee
In the currentless state with $\chi=0,\pi$ one gets from \eqref{sol3b} for the normalization chosen
\be
f_{\infty}^2=\frac{|a|b_n}{a_nb}.
\label{as1}
\ee

The first integrals ${\cal E}_n$ and ${\cal E}$ of the GL equations \eqref{rlambda11}, which are usually useful for 
analyzing the solutions, are
\bal
&{\cal E}_n=\!\left(\dfrac{df(x)}{dx}\right)^2\!\!\!+\dfrac{i^2}{f^2(x)}-f^2(x)-
\dfrac{1}{2}f^4(x), \enspace |x|<l/2, \label{RPhi88gta}\\
&{\cal E}=\left(\dfrac{df(x)}{dx}\right)^2+\dfrac{K_n^2}{K^2}\dfrac{i^2}{f^2(x)}+
\dfrac{|a|K_n}{a_nK}f^2(x) 
- \dfrac{bK_n}{2b_nK}f^4(x),\enspace |x|>l/2,
\label{RPhi88gtb}
\eal
The quantities \eqref{RPhi88gta}, \eqref{RPhi88gtb} are spatially constant, when taken for the solutions to 
\eqref{rlambda11} inside the central electrode and the external leads, respectively. However, the boundary conditions 
\eqref{rlambda30a} generally destroy the conservation of ${\cal E}$ through the interfaces, in contrast to the 
conservation of the supercurrent. Therefore, ${\cal E}_n$ and ${\cal E}$ can substantially differ from each other.

After introducing the function $t(x)=f^2(x)$, the quantity ${\cal E}_n$ for the central electrode can be expressed via 
$t_{\pm}=f_\pm^2$. To this end, one takes $x=(l/2)-0$ in \eqref{RPhi88gta} and excludes the derivatives making use of 
\eqref{jcurbc} and the first equation in \eqref{rlambda30a}:
\be
{\cal E}_n=\biggl[-1+\Bigl(g_{n,\delta}+g_\ell\Bigr)^2\biggr]t_-+g_\ell^2 t_+ 
-2g_\ell\Bigl(g_\ell+g_{n,\delta}\Bigr)\cos\chi \sqrt{t_-t_+}-\dfrac{1}{2}t_-^2\,.
\label{boundcond54b}
\ee

The first GL equation in \eqref{rlambda11} can be analytically solved under the conditions $f_-\sim g_\ell f_+\ll1$ and 
$g_{n,\delta}\alt g_\ell$, which in particular allow one to disregard the cubic term, as compared to the linear one:
\be
\dfrac{d^2f}{dx^2}-\dfrac{i^2}{f^3}-f=0,\quad |x|<l/2.
\label{pap11}
\ee
The solution to \eqref{pap11} can be represented as
\be
f(x)=\sqrt{t_1\cosh^2x-t_2\sinh^2x},
\label{pap12}
\ee
where the parameters $t_1\ge0$ and $t_2\le0$ satisfy
\be
t_1t_2=-i^2, \quad t_1+t_2=-{\cal E}_n.
\label{pap13}
\ee

The order parameter absolute value has the maximums $f_-$ inside the central lead at its end faces $x=\pm l/2$ and the 
minimum $f_1$ at its center $x=0$. As can be confirmed (see \eqref{lsmall132}), under the conditions in question 
the quantity  $f_-$ is of the same order of smallness as $g_\ell f_+$ and, therefore, can satisfy the relation 
$f_-\gg 2g_\ell(g_\ell+g_{n,\delta})f_+$ that reduces the expression \eqref{boundcond54b} for ${\cal E}_n$ to the 
following simplified form
\be
{\cal E}_n=-t_-+g_\ell^2 t_+.
\label{pap14}
\ee

The equalities \eqref{pap12}-\eqref{pap14} and \eqref{jcurbc} lead to the following system of equations
\begin{align}
&t_1t_2=-g^2_{\ell}t_-t_+\sin^2\chi_r, \nonumber\\
&t_1+t_2=t_--g^2_{\ell}t_+, \nonumber\\
& t_-=t_1\cosh^2\frac{l}{2}-t_2\sinh^2\frac{l}{2}.
\label{pap15}
\end{align}
Here $t_-$ and $t_{1,2}$ are on the order of $g^2_{\ell}t_+$, with the higher order terms disregarded.

One finds two solutions to equations \eqref{pap15} that are of the form
\begin{align}
&f_{-,\pm}=g_{\ell}f_+\coth l\biggl[\cos\chi \pm \sqrt{\cos^2\chi-\tanh^2 l}\,\biggr],
\label{lsmall132}\\
&t_{1,\pm}=\frac12 g^2_{\ell} t_+
\biggl\{-2+\biggl(1+\coth^2\frac{l}2\biggr)\cos\chi\biggl[\cos\chi\pm \sqrt{\cos^2\chi-\tanh^2 l}\,
\biggr]\biggr\},
\label{lsmall133}\\
&t_{2,\pm}=\frac12 g^2_{\ell} t_+
\biggl\{-2+\biggl(1+\tanh^2\frac{l}2\biggr)\cos\chi\biggl[\cos\chi\pm \sqrt{\cos^2\chi-\tanh^2 l}\,
\biggr]\biggr\},
\label{lsmall134}
\end{align}
and occur under the condition 
\be
|\chi|\le\chi_{\text{max}}(l)=\arccos\tanh l.
\label{chimax}
\ee
In the same framework $f_\infty$ coincides with its currentless value \eqref{as1} and
$t_+$ is taken in \eqref{lsmall132}-\eqref{lsmall134} in the zeroth order approximation in 
$g_\ell$, i.e., $t_+^{(0)}=\frac{b_n|a|}{ba_n}$ with the normalization used.

As seen in \eqref{lsmall132}, the first solution $f_{-,+}(\chi,l)$ for $f_-$ taken at the end face has its maximum 
at $\chi=0$ and minimum at $\chi=\pm\chi_{\text{max}}(l)$, while the second solution $f_{-,-}(\chi,l)$ for $f_-$ has 
its maximum at $\chi=\chi_{\text{max}}(l)$ and minimum at $\chi=0$:
\begin{align}
&f_{-,+}(0,l)=g_\ell\coth\frac{l}2\left(\frac{b_n|a|}{ba_n}\right)^{1/2}, \label{pap17}\\
&f_{-,-}(0,l)=g_\ell\tanh\frac{l}2\left(\frac{b_n|a|}{ba_n}\right)^{1/2}, \label{pap18}\\
&f_{-,\pm}(\chi_{\text{max}}(l),l)=g_\ell\left(\frac{b_n|a|}{ba_n}\right)^{1/2}.
\label{pap19}
\end{align}
The quantity $f_{-,\pm}(\chi_{\text{max}}(l),l)$ does not depend on $l$ within the approximation used, and the vanishing
value $f_{-,-}(0,l)\to0$ in the limit $l\to0$ agrees with the exact result for the second solution at $\chi=0$ (see 
Supplemental Material in~\cite{Barash2019}).

Substituting the maximal order parameter value \eqref{pap17} in the presumed condition $f_-\ll1$,
one finds that the first solution in  \eqref{pap12}, \eqref{lsmall132}-\eqref{lsmall134}, taken at 
$\cos\chi\sim1$, is justified when the length of the central lead is not too small: 
$l\gg 2g_\ell\left(\frac{b_n|a|}{ba_n}\right)^{1/2}$. For the parameter set used in plotting the figures, one gets 
$l\gg 0.02$. The first solution at sufficiently  small $l$ has been considered in Supplemental 
Material in~\cite{Barash2019}. One notes that the results of this section can be applied at any $l$ to the second 
solution, as well as to the first one in a vicinity of $\chi=\chi_{\text{max}}(l)$. 

For the absolute order parameter value at the center of the normal metal lead  $f(x=0,\chi,l)\equiv f_{1}(\chi,l)$, one 
finds from \eqref{pap12} and \eqref{lsmall133}:
\begin{align}
&f_{1,+}(0,l)=\dfrac{g_\ell}{\sinh\frac{l}2}\left(\frac{b_n|a|}{ba_n}\right)^{1/2}, \label{pap20}\\
&f_{1,-}(0,l)=0, \label{pap21}\\
&f_{1,\pm}(\chi_{\text{max}}(l),l)=\dfrac{g_\ell}{\sqrt{\cosh l}}\left(\frac{b_n|a|}{ba_n}\right)^{1/2}.
\label{pap22}
\end{align}

The vanishing second solution in \eqref{pap21} at the center of the lead at $\chi=0$ makes possible phase-slip processes
at $\phi=\pi$, and at arbitrary $l$, in agreement with earlier results~\cite{Fink1976,Giazotto2017}. The corresponding
phase incursion is $\varphi=\pi$ and the supercurrent vanishes under such a condition. As seen in \eqref{jcurbc}, the 
phase incursion can differ from zero in the limit $i\to0$ only if the order parameter $f(x)\ge0$ takes its minimum value
$f=0$ somewhere inside the central electrode. This is the case at $x=0$.

One obtains from \eqref{jcurbc} and the solutions \eqref{pap12}, \eqref{lsmall132} - \eqref{lsmall134}
the following expression for the phase incursion $\varphi$:
\be
\varphi_\pm(\chi,l)= \sgn(\sin\chi)\arccos\biggl[\cosh l\biggl(\sin^2\chi\pm 
\cos\chi\sqrt{\cos^2\chi-\tanh^2 l}\biggr)\biggr].
\label{lsmall158}
\ee
One gets $\varphi_+(0,l)=0$ from \eqref{lsmall158} for the first solution and $\varphi_-(0,l)=\pi$
for the second one and $\varphi_\pm(\chi_{\text{max}}(l),l)=\arccos(1/\cosh l)$, where 
$\chi_{\text{max}}(l)$ is defined in \eqref{chimax}. Therefore, there are two possible values $0$ and $\pi$ of the phase
incursion $\varphi$ in currentless states of the double junctions in question, as have been earlier identified in 
Refs.~\cite{VolkovA1971} and \cite{Fink1976}.

For the external phase difference, which satisfies the relation $\phi(\chi,l)=2\chi+\varphi_\pm(\chi,l)$,
one finds
\be
\sin\phi_\pm(\chi,l)=
\Biggl[\cos\chi\pm \sqrt{\cos^2\chi-\tanh^2 l}\Biggr]\cosh l\sin\chi.
\label{lsmall168}
\ee
For $\phi_*(l)\equiv\phi(\chi_{\text{max}}(l),l)$ one gets from \eqref{lsmall168}:
\be
\phi_*(l)=\dfrac{\pi}{2}+\arcsin(1/\cosh l).
\label{pap30}
\ee

Thus, for the double tunnel junction considered, it follows from \eqref{jcurbc}, \eqref{lsmall132} and \eqref{lsmall168}
the conventional sinusoidal current-phase relation with respect to the external phase difference:
\be
i=\dfrac{g_\ell^2}{\sinh l}t_+\sin\phi.
\label{pap31}
\ee 

Deviations of the dashed curves from solid ones in Figs.~\ref{fig:chi-varphi-phi-l} - \ref{fig:cur-chi-phi-l} 
demonstrate a reasonable accuracy of the analytical results of this subsection with respect to the corresponding results
of exact numerical calculations. Although, the accuracy of the phase relations \eqref{lsmall158}-\eqref{pap30} is 
noticeably better as compared with the expressions \eqref{pap12}, \eqref{lsmall132}-\eqref{lsmall134} obtained for the 
absolute value of the order parameter within the same approximation. One notes that the coupling constants have canceled
out in \eqref{lsmall158}-\eqref{pap30}, as distinct from \eqref{lsmall132}-\eqref{lsmall134}.

\section{Internal Phase Differences in Symmetric \mbox{SISIS} Double Junctions}
\label{sec: sisis}

\subsection{Preliminary Remarks}

As was demonstrated in the preceding section for \mbox{SINIS} double junctions, the Josephson coupling associated with 
the phase-dependent bilinear part of free energy $-2g_J\left|\Psi_-\right|\left|\Psi_+\right|\cos\chi$ (see  
\eqref{Jos1}) is responsible for the supercurrent on condition that $g_\ell\cos\chi>0$. This is necessary for 
a nonzero order parameter $\left|\Psi_-\right|$ on the normal metal interface side to be induced in the presence of 
$\left|\Psi_+\right|$ on the opposite superconducting side of a thin interface. As a result, for superconductivity to be
proximity-induced over zero junctions ($g_J>0$) in the normal metal electrode, the internal phase difference 
$\chi$ must be confined within a substantially reduced range $|\chi|\le\chi_{\text{max}}(l)<\frac{\pi}{2}$ \, 
(modulo $2\pi$). Outside of that range the Josephson coupling prevents superconductivity from showing up in the normal 
metal lead, within the model discussed. The quantity $\chi_{\text{max}}(l)$ is less than $\frac{\pi}{2}$ due to the 
effect of the current-induced phase incursion along the central lead. In the \mbox{SISIS} systems with the 
central-electrode length of a mesoscopic size, the effect of the phase incursion is usually negligibly small except for 
an immediate vicinity of $|\chi|=\pi/2$.

Unlike \mbox{SINIS} double junctions, all three electrodes in \mbox{SISIS} heterostructures possess identical
original superconducting properties. Proximity and pair-breaking effects are not necessary in the latter case for
producing the supercurrent itself, although they can noticeably modify its properties under certain conditions. For zero
junctions, the phase-dependent part of the Josephson interface free energy $-2g_J\left|\Psi_-\right|\left|\Psi_+\right|
\cos\chi$ shifts the energies of the states with $|\chi|<\pi/2$ and $\pi/2<|\chi|<\pi$ down and up, respectively. In 
other words, the Josephson coupling is responsible for a superconductivity enhancement under the condition 
$|\chi|<\pi/2$, while for $\pi/2<|\chi|<\pi$ it has pair breaking character. This results in the presence of lower and 
higher energy modes in the system. 

When the central electrode's length $L$ substantially exceeds the superconductor coherence length, the internal phase 
difference $\chi$ can generally take any value in \mbox{SISIS} superconducting double junctions. However, with 
decreasing length $L$, a difference between the energies of the two types of modes increases, and the pair breaking of 
the Josephson origin can play a crucial role in destroying the higher energy modes, when $L$ is less than a phase 
dependent critical value~\cite{Barash2018}. This results in a reduced length-dependent range of variations of $\chi$. In 
particular, as will be seen below, the lower energy state with $\chi_{1}=\chi_{2}=0$ occurs at an arbitrary $L$, while 
the higher energy equilibrium state with $\chi_{1}=\chi_{2}=\pi$ exists only for $L$ exceeding a critical length 
$L_{\pi}$. The formation of the proximity-reduced range of the internal phase difference in \mbox{SISIS} double junctions
also depends on the Josephson coupling strength. In the limit of vanishing strength, the range agrees with the known 
result, which states that superconductivity is completely destroyed in a metal film sandwiched between two impenetrable
pair breaking walls when the film's thickness $L$ is less than its critical (phase-independent) 
value~\cite{Ginzburg1958,ROZaitsev1965,Ginzburg1993,Barash2017}. 

On the other hand, the Josephson current is, generally speaking, not uniquely defined at a fixed phase difference 
$\phi$ between the external leads, when a sufficiently wide range of $\chi_{1,2}$ is allowed. Consider, for example, the 
relation $\phi=\chi_1+\chi_2$ assuming the phase incursion over the central lead to be negligibly small. For symmetric
double junctions one can take $\chi_1=\chi_2+2\pi n$ with integer $n$, that results in $\chi_1=\frac{\phi}2+\pi n$. 
Therefore, after switching from $\chi_{1,2}$ to $\phi$, the $2\pi$-periodic current-phase relation $j(\chi_1)$ for a single 
junction is transformed into two different $4\pi$-periodic modes $j(\frac{\phi}2)$ and $j(\frac{\phi}2+\pi)$, with 
respect to $\phi$. The change of $\phi$ by $2\pi$, caused by the variations of both $\chi_{1,2}$ by $\pi$, results in 
the supercurrent sign change for each separate mode. Although, if the same change $\phi\to \phi+2\pi$ occurs when only 
one of $\chi_{1,2}$ varies by $2\pi$, the current remains unchanged. Thus, the supercurrent is a double-valued function 
of $\phi$ that can be verified, when $\chi_{1,2}$ are independently controlled in experiments.

However, if the external phase difference $\phi$ is experimentally controlled, $\chi_{1,2}$ can take on the most 
preferable equilibrium values. The first mode $j(\frac{\phi}2)$ is energetically favorable under the condition
$(4n-1)\pi\le\phi\le(4n+1)\pi$, while the second mode $j(\frac{\phi}2+\pi)$ is favorable at 
$(4n+1)\pi\le\phi\le(4n+3)\pi$. Therefore, the two $4\pi$-periodic states get interchanged when the external phase 
$\phi$ changes by $2\pi$, if disregarding possible ``undercooling'' or ``overheating'' of the states during the 
transition. Such a regime of interchanging modes is characterized by a $2\pi$-periodic sawtooth-like current-phase 
relation with discontinuities at $\phi=(2n+1)\pi$~\cite{Luca2009,Linder2017,Barash2018}, as shown below in the curve 1 
in the right panel of Fig.~\ref{fig:cur}. It is a characteristic feature of conventional superconducting double junctions, 
considered both in this and the preceding sections, that a potential $4\pi$-periodic current-phase dependence, related 
to the external phase difference $\phi$, gets, for one reason or another, broken and transformed into a $2\pi$-periodic
behavior. This differs clearly from the real $4\pi$-periodic supercurrent through Josephson junctions involving Majorana
fermions~\cite{Kitaev2001,Kane2009,Beenakker2013,Alicea2012,Molenkamp2016,Molenkamp2017}.

A sharp change of the supercurrent, that takes place in this regime in immediate vicinities of $\phi_n=(2n+1)\pi$, can 
occur continuously and involve the current-carrying asymmetric states. In the tunneling limit, for example, the symmetry
$j(\pi-\chi)=j(\chi)$ allows one to get the value $\phi=\chi_1+\chi_2=\pi$ with $\chi_2=\pi-\chi_1$ at all possible 
$\chi_1$ and, therefore, at any admissible value of the supercurrent $|j|\le j_c$. For sufficiently small central 
electrode length, the proximity effects reduce the range of $\chi_{1,2}$ and the order parameter values in the central
lead. As a result, the regime of interchanging modes is destroyed and the conventional smooth current-phase relation 
$j(\phi)$ is restored at all $\phi$ in this limit~\cite{Barash2018}. The abrupt change, produced by a competition of the
doubled condensate states at given $\phi$, can also be partially smeared out by fluctuations, small junction
asymmetries etc~\cite{Linder2017}. 

In the limit of small $L$, the double Josephson junction can be considered as an effective single junction with a thin 
interface involving the central region. Although this is a specific interface, where only a sequential tunneling takes 
place, while a direct one is forbidden within the model in question, the single-junction regular phase dependence 
$j(\phi)$ at all $\phi$ occurs in this limit. The experimental data support this issue~\cite{Blamire1994}. The 
microscopic studies revealed the conventional single junction behavior at a very small $L$, but without taking the 
regime of interchanging modes at the larger $L$ into 
account~\cite{Kupriyanov1988,Kupriyanov1999,Golubov2000,Ishikawa2001}. The proximity effects at $L\ll \xi$ have been 
known to be resposible for a pronounced order parameter's phase-dependence in the central lead.

A theory of the proximity-influenced regime of interchanging modes and of the behavior of internal phase differences in
symmetric \mbox{SISIS} double Josephson junctions considered below, was developed in Ref.~\cite{Barash2018} within 
the GL approach based on the interface boundary conditions discussed in Sec.~\ref{sec: bc}.

\subsection{Model and Basic Equations}

\begin{figure}[t]
\centering
\includegraphics*[width=.4\textwidth,clip=true]{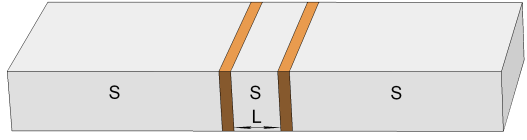}
\caption{Schematic diagram of the double \mbox{SISIS} junction. Adopted from Ref.~\cite{Barash2018}.} \label{fig: dj}
\end{figure}

Consider a symmetric \mbox{SISIS} double junction, shown in Fig.\,\ref{fig: dj}, that contains two identical thin
interfaces at a distance $L$, connected by the central superconducting electrode made of the same superconducting 
material as the external leads. Similar to Secs.~\ref{sec: bc} - \ref{sec: sinis}, a comparatively small interface
thickness is assumed for defining it to be zero within the GL theory. The length of the external leads is supposed to be
large, significantly exceeding the coherence length $\xi(T)$. The one-dimensional spatial dependence of the order 
parameter considered below occurs, for example, when the transverse dimensions of all three electrodes are significantly
less than $\xi(T)$ as well as the magnetic penetration depth. 

The system's free energy incorporates the interface and bulk contributions
${\cal F}=\sum{\cal F}_{p}+{\cal F}^{\text{int}}_{\frac{L}2}+{\cal F}^{\text{int}}_{-\frac{L}2}$, where $p=1,2$
correspond to the external electrodes and $p=3$ refers to the central lead. The bulk free energies per unit area of the 
cross section are
\be
\!\!{\cal F}_p\!=\!\!\!\int\limits_{{\cal C}_p}\!\!dX\!
\left[\!K\left|\dfrac{d}{dX}\Psi(X)\right|^2\!\!\!+a\left|\Psi(X)\right|^2\!\!+\dfrac{b}{2}\left|\Psi(X)\right|^4\right].
\label{Fb1}
\ee
The integration periods ${\cal C}_p$ for $p=1,2,3$ are taken to be $(-\infty,-L/2)$, $(L/2,\infty)$ and $(-L/2,L/2)$, 
respectively.

For each of two interfaces placed at $X=\pm L/2$, the interfacial free energy per unit area is, in accordance with 
\eqref{Jos5},  
\begin{equation}
{\cal F}^{\text{int}}_{\pm\frac{L}2}=g_{J}\left|\Psi_{\pm\frac{L}{2}+}-\Psi_{\pm\frac{L}{2}-}\right|^2\!\!
+g\left(\left|\Psi_{\pm\frac{L}{2}+}\right|^2\!\!+\left|\Psi_{\pm\frac{L}{2}-}\right|^2\right).
\label{fint1}
\end{equation}
For $0$-junctions considered below the Josephson coupling constant $g_J>0$. The parameter $g>0$ describes the strength 
of the surface pair breaking.
 
The GL equation, applied to the normalized absolute value of the superconductor order parameter 
$\Psi=(|a|/b)^{1/2}fe^{\mathtt{i}\alpha}$, takes the form
\begin{equation}
\dfrac{d^2f}{dx^2}-\dfrac{i^2}{f^3}+f-f^3=0,
\label{gleq2}
\end{equation}
where $x=X/\xi(T)$, $\xi(T)=(K/|a|)^{1/2}$ and the dimensionless current density is 
$i=\frac{2}{3\sqrt{3}}(j\big/j_{\text{dp}})$, where 
$j_{\text{dp}}=\bigl(8|e||a|^{3/2}K^{1/2}\bigr)\big/\bigl(3\sqrt{3}\hbar b\bigr)$ is the 
depairing current deep inside the external superconducting leads. 

The boundary conditions, that follow from \eqref{Fb1} and \eqref{fint1}, are in agreement with \eqref{Jos6}.
After introducing the dimensionless real quantities, the conditions at $x=l/2$, where  $l=L/\xi(T)$, can be written as
\begin{align}
&\left(\dfrac{df}{dx}\right)_{l/2\pm0}\!\!=
\pm\Bigl(g_\delta+g_\ell\Bigr)f_{l/2\pm0}\mp g_\ell\cos\chi f_{l/2\mp0},
\label{bc1} \\
&i=-\,f^2\left(\dfrac{d\alpha}{dx}+\dfrac{2\pi\xi(T)}{\Phi_0}A\right)
=g_{\ell}f_{l/2-0}f_{l/2+0}\sin\chi .
\label{joscurr1}
\end{align}
Here $\chi=\alpha\left(\frac{l}{2}-0\right)-\alpha\left(\frac{l}{2}+0\right)$, $\Phi_0=\frac{\pi\hbar c}{|e|}$
and the dimensionless coupling constants are introduced $g_\ell=g_J\xi(T)/K$, $g_{\delta}=g\xi(T)/K$.

As this follows from the boundary conditions \eqref{bc1} and the conservation of the supercurrent \eqref{joscurr1},
the order parameter values $f_{l/2\pm0}$ on opposite sides of an interface between identical superconductors can 
generally differ from one another. This is usually not the case in a single symmetric Josephson junction, where $f(x)$ 
is continuous across a thin interface. However, the condensate density can be noticeably weakened by pair breaking 
interfacial effects on both ends of the central lead of mesoscopic length of the double junction. The resulting phase
dependent jump $f_{l/2+0}-f_{l/2-0}>0$ plays an important role allowing superconductivity to survive in the central lead
with a small length $l\ll1$ despite the interface pair breaking. The fallacious continuity of $f(x)$ across thin 
interfaces in double Josephson junctions is a characteristic feature of earlier theories that used the flawed boundary 
conditions for the order parameter within the GL approach~\cite{Baratoff1975,Kao1977,SolsZapata1996}. As follows from 
Subsec.~\ref{subsec: micro}, those models disagree with the microscopic results near $T_c$.

A number of solutions to equation \eqref{gleq2} were found that satisfy the asymptotic conditions in the external 
electrodes and the boundary conditions at $x=\pm l/2$ (see Appendix A in Ref.~\cite{Barash2018} for details). The 
solutions that possess the extrema only at $x=0,\,\pm l/2,\, \pm\infty$, or, when possible, only at 
$x=\pm l/2,\,\pm\infty$ are assumed to have preferable energies. The symmetric solutions $f(x)=f(-x)$ with the internal
phase differences $\chi_1=\chi_2+2\pi n=\chi$, exist in most cases considered below, with the exception of close 
vicinities of $\phi_n=(2n+1)\pi$, where the asymmetric behavior can take place.

\subsection{Currentless States at $\chi=0,\,\pi$}

\begin{figure}[t]
\begin{center}
\begin{minipage}{.49\columnwidth}
\includegraphics*[width=.61\columnwidth,clip=true]{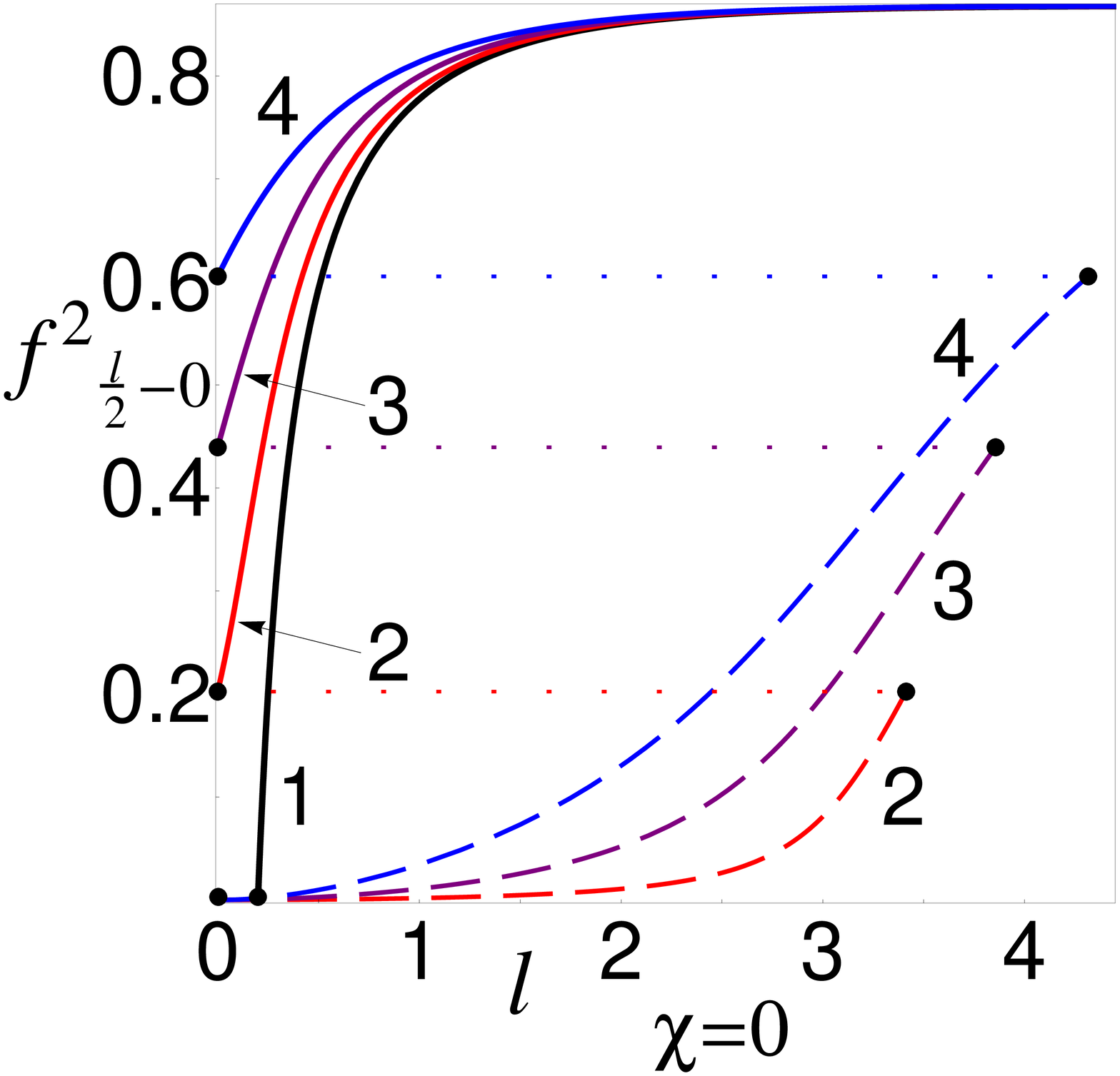}
\end{minipage}
\begin{minipage}{.49\columnwidth}
\includegraphics*[width=.64\columnwidth,clip=true]{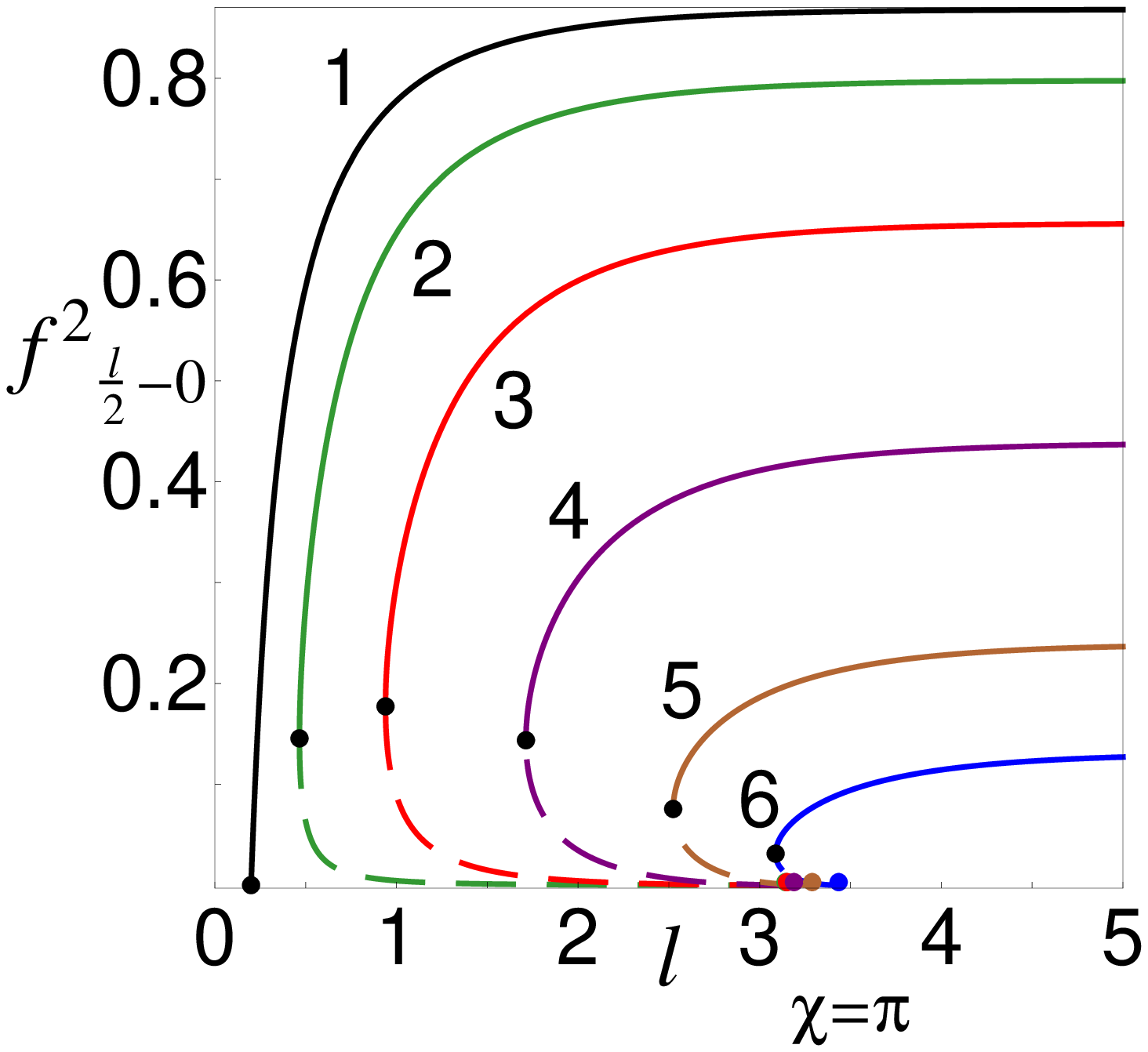}
\end{minipage}
\end{center}
\caption{$f^2_{l/2-0}$ as a function of $l$ 
at $\chi=0$ (left panel) and $\chi=\pi$ (right panel). 
Solid curves correspond to the energetically preferable states. 
{\it Left panel}: $\chi=0$,\, $g_\delta=0.1$ and  \,\,
(1)\, $g_{\ell}=0$\,\,
(2)\, $g_{\ell}=0.1$\,\,
(3)\, $g_{\ell}=0.3$,\,\, and \,\,
(4)\, $g_{\ell}=0.8$.
{\it Right panel}:
$\chi=\pi$,\, $g_\delta=0.1$ and \,\,
(1)\, $g_{\ell}=0$\,\,
(2)\, $g_{\ell}=0.03$\,\,
(3)\, $g_{\ell}=0.1$,\,\, 
(4)\, $g_{\ell}=0.25$,\,\, 
(5)\, $g_{\ell}=0.5$,\,\,and \,\,
(6)\, $g_\ell=0.8$. Adopted from Ref.~\cite{Barash2018}.}
\label{fig:rminchi}
\end{figure}

The double junction's states can be described analytically in the absence of the supercurrent, i.e., at 
$\chi=\pi n$~\cite{Barash2018}. Fig.\,\ref{fig:rminchi} shows the dependence of the quantity 
$f^2_{l/2-0}(\chi,g_\ell,g_\delta)$, taken at the end side of the central lead, on its dimensionless length $l$ at 
$\chi=0$ (the left panel) and $\chi=\pi$ (the right panel), calculated with $g_\delta=0.1$ at various $g_\ell$. 
The solid curves correspond to energetically preferable solutions, while the dashed curves describe metastable states.

The curves 1 in both panels in Fig.\,\ref{fig:rminchi} are identical due to vanishing dependence on the phase difference
for an impenetrable wall ($g_\ell=0$). As known, superconductivity is destroyed in the sample placed between two 
impenetrable pair breaking walls with decreasing distance $L$ between them, and the transition to the normal metal state
occurs at $L=2\xi(T)\arctan g_\delta$~\cite{ROZaitsev1965,Ginzburg1993,Barash2017}. For $g_\delta=0.1$ one gets from 
here $l=0.199$. However, as seen in the solid curves 2-4 in the left panel of Fig.\,\ref{fig:rminchi}, the nonzero
minimum of the quantity $f^2_{l/2-0}$ is at $l=0$ in the presence of a finite interface transparency ($g_\ell>0$), but 
not in the limit $g_\ell\to0$ at $g_\delta\ne0$. Therefore, despite the surface pair breaking, the finite interface
transparency allows the superconducting state at $\chi=0$ to exist in the central electrode at any small value of its 
length.

According to the boundary condition  $(df/dx)_{(l/2)-0}=-(g_\delta+g_\ell)f_{(l/2)-0}+g_\ell f_{(l/2)+0}$ taken at 
$\chi=0$, the order parameter derivative is negative under the condition 
$f_{(l/2)-0}>g_\ell f_{(l/2)+0}/(g_\delta+g_\ell)$ and, therefore, $f(x)$ increases, when $x>0$ goes down to $x=0$.
The corresponding solutions are shown in the solid curves 2 - 4 in the left panel of Fig.~\ref{fig:rminchi}. The 
derivative $(df/dx)_{(l/2)-0}$ vanishes at the end face of the central lead, if the equality $f_{(l/2)-0}=
g_\ell f_{(l/2)+0}/(g_\delta+g_\ell)$ holds. The dashed curves in the left panel of Fig.\,\ref{fig:rminchi} correspond
to the solution of a different type, for which the order parameter derivative is positive $(df/dx)_{(l/2)-0}>0$ and 
$f(x)$ decreases with decreasing $x$ inside the central lead vanishing at $x=0$. This is a metastable solution that 
takes smaller order parameter values and satisfies the relation $f_{(l/2)-0}<g_\ell f_{(l/2)+0}/(g_\delta+g_\ell)$.

As distinct from the case $\chi=0$, both terms on the right-hand side of the boundary conditions \eqref{bc1} have 
negative sign at $\chi=\pi$, as well as at all $\pi/2<|\chi|\le\pi$. For this reason, the sign of the order 
parameter derivative at $\chi=\pi$ is negative in the central electrode irrespective of the relation between 
$f_{(l/2)-0}$ and $f_{(l/2)+0}$. In this case the condensate density at $x>0$ always decreases the nearer one gets to 
the interface. Under such conditions, the state with $\chi=\pi$ exists only if the central electrode's length $l$ 
exceeds the critical value $l_{\pi}(g_\ell,g_\delta)$. However, in contrast to what takes place at $g_\ell\equiv0$, a 
disappearance of the equilibrium state with $\chi=\pi$ at $l<l_{\pi}(g_\ell,g_\delta)$ and $g_\ell\ne0$ is not 
associated with a system's transition to the normal metal state.

Since the system involves distant superconducting regions of the external electrodes, the transition to the 
normal metal state cannot be induced by the interfacial pair breaking which is confined by the scale $\alt\xi(T)$. Also,
the central electrode cannot individually be in the normal metal state ($f(x)=0$ at $|x|<l/2-0$) once $g_\ell\ne0$. In 
this case the boundary condition $(df/dx)_{(l/2)-0}=-(g_\delta+g_\ell)f_{(l/2)-0}-g_\ell f_{(l/2)+0}$ at $x=l/2-0$ and 
$\chi=\pi$ would result in $f_{(l/2)+0}=0$. One would also obtain $(df/dx)_{(l/2)+0}=0$ from the boundary condition on 
the opposite side of the interface. This would mean the normal metal state of the external leads, which is not possible 
as stated above. Therefore, superconductivity does exist under the conditions $l<l_{\pi}(g_\ell,g_\delta)$ and 
$g_\ell\ne0$ due to the proximity to the external superconducting leads, while the value $\chi=\pi$ cannot be the 
equilibrium value of $\chi$. Similar effects, that reduce the range of variation of $\chi$, take place for all 
$\pi/2<|\chi|\le\pi$ with the phase-dependent critical length $l_{\chi}(g_\ell,g_\delta)$.

The dashed curves in the right panel of Fig.\,\ref{fig:rminchi} describe metastable solutions at $\chi=\pi$. They appear
within the range $l_{\pi}(g_\ell,g_\delta)<l<l_{ps}(g_\ell,g_\delta)$ being of the same type as the energetically 
preferable solutions. At $l=l_{ps}(g_\ell,g_\delta)$, the metastable phase-slip centers, where $f_{\pm(l_{ps}/2-0)}=0$,
appear at the central lead's end interfaces. The length $l_{ps}$ takes on its minimum value 
$l_{ps}(g_\ell\to0,g_\delta)=\pi$ in the tunneling limit. The points with coordinates $l=l_{ps}(g_\ell,g_\delta)$ and 
$f_{l/2-0}=0$ are marked in the right panel of Fig.\,\ref{fig:rminchi}.

The numerical study of the solutions shows that the left and right panels of Fig.\,\ref{fig:rminchi} depict
the two basic types of mapping functions $f^2_{l/2-0}$. One type is transformed into another with $\chi$ 
at some distance below $\chi=\pi/2$.

\subsection{Central Electrode with a Small Length $l\ll1$}

In the limit of small central electrode's length $l$ at a given phase difference $\chi$, the order parameter absolute 
values on opposite interface sides are linked by a relation that follows from the boundary conditions and corresponds to 
approximately vanishing order parameter derivative in the central region~\cite{Barash2018}. As seen from the boundary 
conditions \eqref{bc1}, the derivative $\left({df}/{dx}\right)_{l/2-0}$ vanishes, if, similar to \eqref{lto0}, 
$\cos\chi>0$ and the order parameter absolute values satisfy the relation:
\be
f_{l/2-0}=\dfrac{g_\ell\cos\chi}{g_\ell+g_\delta}f_{l/2+0}.
\label{smalll1}
\ee  

This demonstrates that, in the limit of small length $l$, a superconducting state survives in the central electrode 
under the condition $\cos\chi>0$ ($g_\ell>0$). The pair breaking effects of the Josephson origin prohibit in this  
limit the equilibrium values of internal phase differences with $\cos\chi<0$.
 
Internal phase differences that satisfy the condition $\cos\chi>0$, form allowed bands $-\pi/2+2\pi n<\chi<\pi/2+2\pi n$
separated by the forbidden gaps. After switching over to the external phase difference $\phi$ and disregarding the phase
incursion over the central lead, one obtains the same bands for the argument $\frac{\phi}2$ of the first mode and
for the argument $\frac{\phi}2+\pi$ of the second mode. Therefore, the functions $\cos\chi$ and $\sin\chi$ correspond 
here to $\cos\frac{\phi}2$ and $\sin\frac{\phi}2$, if $(4n-1)\pi\le\phi\le(4n+1)\pi$, and to
$\cos(\frac{\phi}2+\pi)=-\cos\frac{\phi}2$ and $-\sin\frac{\phi}2$ in the case $(4n+1)\pi\le\phi\le(4n+3)\pi$. 
Thus, one obtains from \eqref{smalll1} at any value of $\phi$
\begin{align}
&f_{l/2-0}=\frac{g_\ell|\cos\frac{\phi}2|}{g_\delta+g_\ell}f_{l/2+0}, \label{rel1}\\
i=g_\ell^{\text{eff}}&f^2_{l/2+0}\sin\phi, \quad g_\ell^{\text{eff}}=\dfrac{g_\ell^2}{2(g_\delta+g_\ell)},
\label{joscurr2}
\end{align}
where the right hand side in \eqref{joscurr1} has been used in \eqref{joscurr2}.

Thus, in the limit $l\ll1$, the allowed bands from both modes for the external phase difference $\phi$ are adjoined with
no overlapping portions and with no gaps. This leads to the single-valued dependence on $\phi$ of the quantities 
considered, at all values of $\phi$. However, while the higher energy mode is completely destroyed in the limit
of very small $l$ due to the proximity reduced range of the internal phase differences, it is present to the full 
extent at comparatively large $l$. The total annihilation of the condensate states' doubling at any given $\phi$ and the
expression \eqref{joscurr2} for the supercurrent demonstrate that the double junction behavior in the limit $l\to0$ is
transformed to that of a symmetric single junction with the effective Josephson coupling constant $g_\ell^{\text{eff}}$.

The boundary condition for a single symmetric Josephson junction~\cite{Barash2012} follows from \eqref{bc1} under the 
condition $f_{l/2-0}=f_{l/2+0}$ that allows one to exclude the order parameter taken at the internal interface side: 
\be
\left(\dfrac{df}{dx}\right)_{l/2+0}\!\!=\Bigl(g_\delta+2g_\ell\sin^2\frac{\chi}{2}\Bigr)f_{l/2+0}.
\label{bc1s}
\ee

For the double junctions with $l\ll1$, one can exclude, based on \eqref{rel1}, the internal value $f_{l/2-0}$ in 
\eqref{bc1} in favor of $f_{l/2+0}$ and obtain an approximate boundary condition of the same form as \eqref{bc1s}, but 
with the effective Josephson coupling constant $g_\ell^{\text{eff}}$, defined in \eqref{joscurr2}, and the
effective interfacial pair breaking parameter 
\be
g_\delta^{\text{eff}}=\frac{g_\delta(g_\delta+2g_{\ell})}{g_\delta+g_{\ell}}. 
\label{gdeltaeff}
\ee
 
In the limit $l\ll1$, the proximity effects play a crucial role in establishing the conventional sinusoidal 
current-phase relation \eqref{joscurr2} in the \mbox{SISIS} double junction, if the sequential tunneling dominates
the direct one in the system. This is the proximity-induced phase dependent factor $|\cos\frac{\phi}2|$ on the 
right-hand side of \eqref{rel1}, which is responsible in this case for the supercurrent \eqref{joscurr2} to decrease 
with increasing $\phi$ at $\pi/2\le\phi\le\pi$. As discussed in Subsec.~\ref{subsec: micro},
in tunnel junctions $g_\ell\propto{\cal D}$, where $\cal D$ is the interface transmission coefficient. Therefore one
obtains from \eqref{joscurr2} $g_\ell^{\text{eff}}\propto {\cal D}^2$ under the condition $g_\delta\gg g_\ell$, and 
$g_\ell\propto {\cal D}$ in the opposite limit $g_\delta\ll g_\ell$, in agreement with the earlier microscopic 
results~\cite{Kupriyanov1988,Kupriyanov1999,Golubov2000}. 

A pronounced suppression of the quantity $f_{l/2-0}$ in a close vicinity of $\phi=\pi$, that follows from \eqref{rel1}, 
and the spatial uniformity of the supercurrent \eqref{joscurr1} result in a large gradient of the order-parameter phase.
The numerical data show that a noticeable phase incursion over the central lead arises for this reason even at very 
small $l$ and violates near $\phi=\pi$ the applicability of the approximation used. After switching over to $\chi$, this
results in no discernible modifications in \eqref{rel1} and \eqref{joscurr2}, but the range of $\chi$ diminishes so that
only values at a distance below $\chi=\pi/2$ are permitted at small~$l$:\,$\chi(\phi)<\pi/2$ at $\phi=\pi$.

\begin{figure}[t]
\begin{center}
\begin{minipage}{.49\columnwidth}
\includegraphics*[width=.7\columnwidth,clip=true]{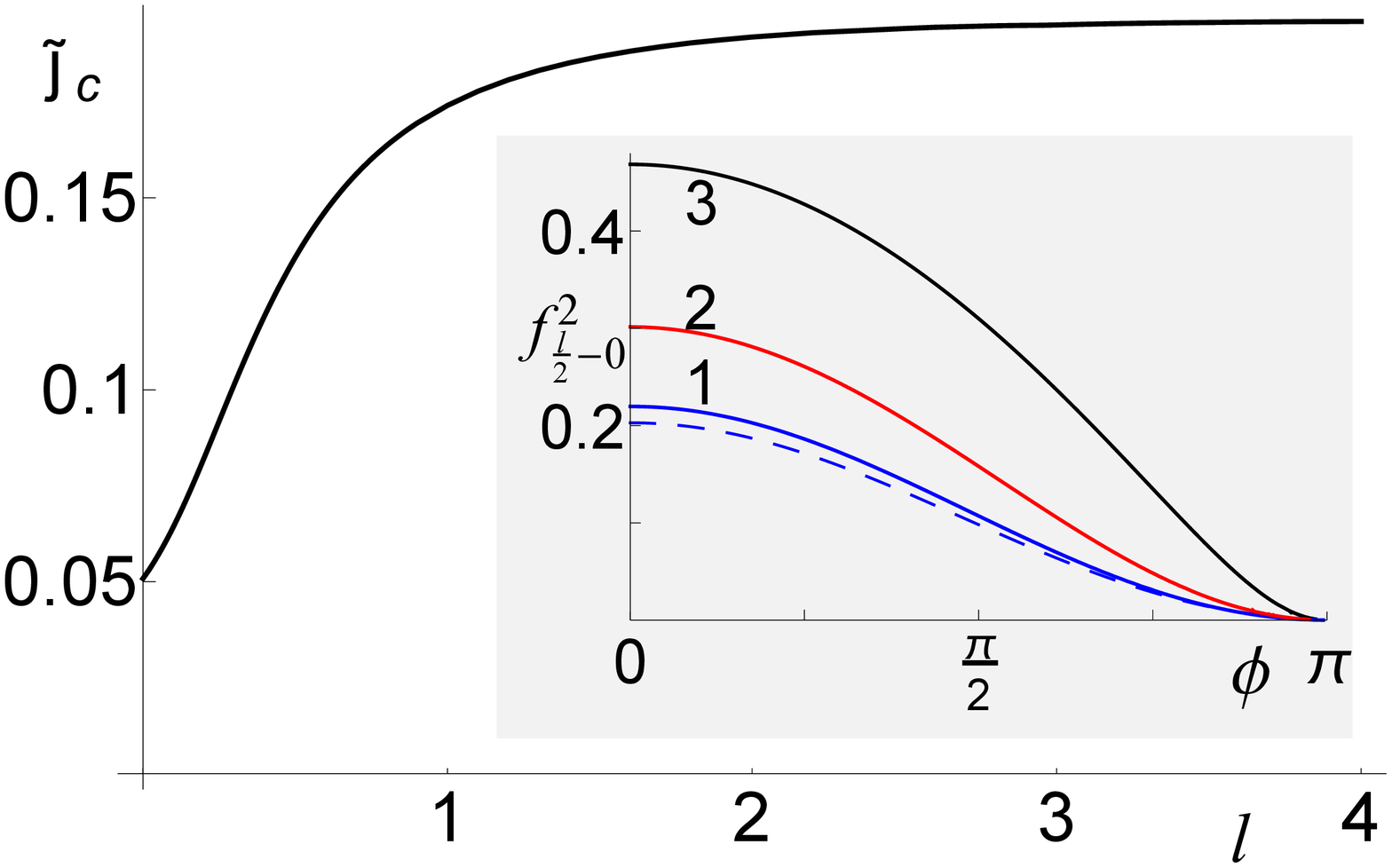}
\end{minipage}
\begin{minipage}{.49\columnwidth}
\includegraphics*[width=.74\columnwidth,clip=true]{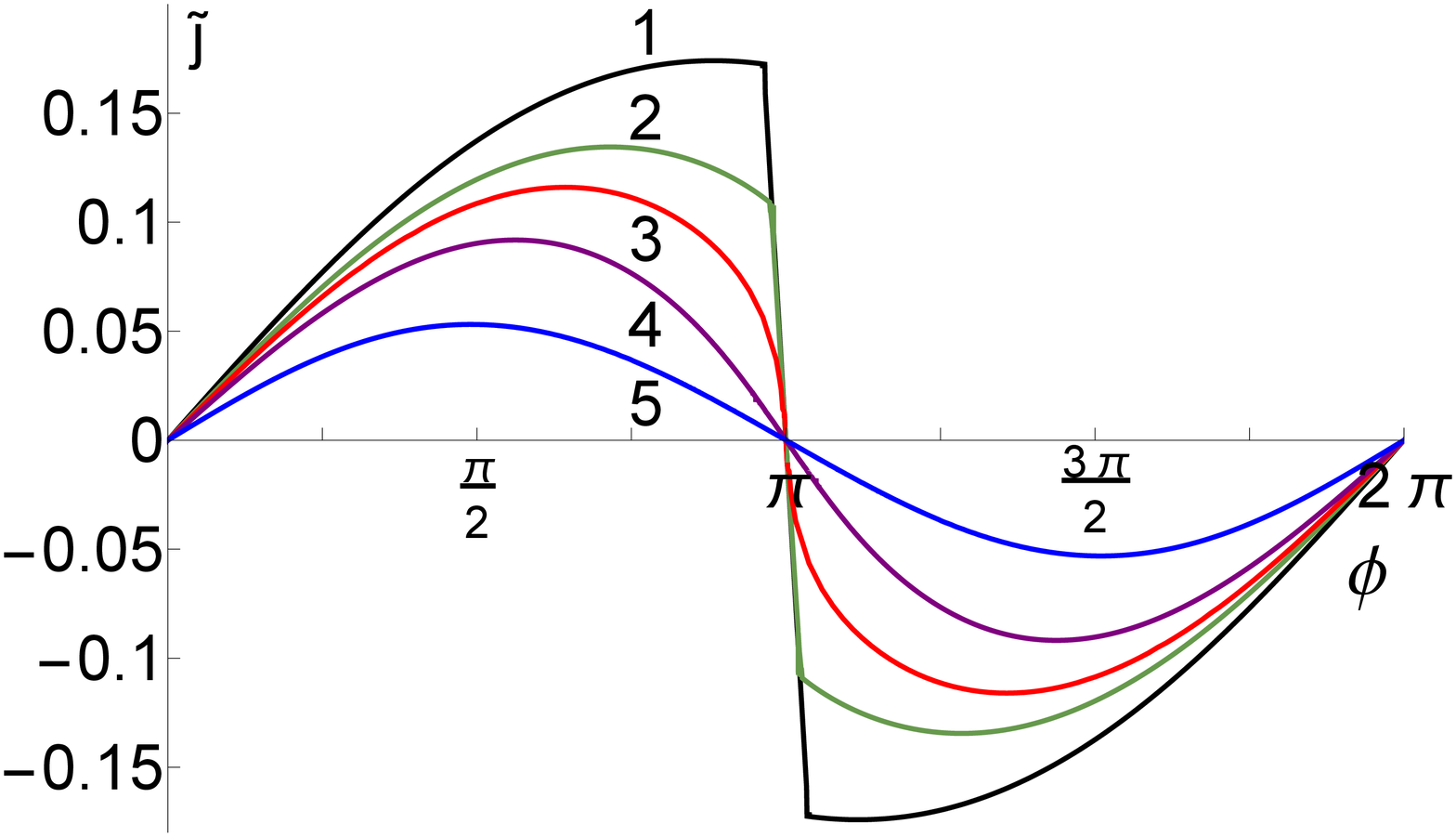}
\end{minipage}
\end{center}
\caption{Left panel: Critical current as a function of $l$ at $g_{\ell}=0.1$, $g_{\delta}=0.1$.\,Inset: 
The quantity $t_{l/2-0}(\phi)$ at (1)\, $l=0.02$\,\, (2)\, $l=0.1$\,\, (3)\, $l=0.25$. 
Right panel: Current-phase relations $\tilde\jmath(\phi)$ taken for $g_{\ell}=0.1$,\, $g_\delta=0.1$ and \,\, (1)\, 
$l=1$\,\, (2)\, $l=0.5$\,\, (3)\, $l=0.38$\,\, (4)\, $l=0.25$\,\, (5)\, $l=0.02$. Adopted from Ref.~\cite{Barash2018}.}
\label{fig:cur}
\end{figure}

The inset in the left panel of Fig.\,\ref{fig:cur} shows the phase dependence of order parameter squared $f^2_{l/2-0}$
at the end face of the central lead. The solid curves 1 - 3 correspond to the numerical results. The dashed 
curve depicts the analytical results, which follow from the right-hand side of \eqref{rel1} at $l=0.02$ and deviates 
only by a few percent from the curve~1. However, the dashed curves at $l=0.1$ and $l=0.25$ (not shown) almost coincide 
with the one presented for $l=0.02$ and, therefore, substantially deviate from the solid curves 2 and 3. In other words,
the relation \eqref{rel1}, which is justified at $l=0.02$ for the chosen set of parameters, is, with increasing $l$,
in contradiction with the careful numerical calculations already at $l=0.1$ and $l=0.25$. The origin of the discrepancy
is that the right-hand side of \eqref{rel1}, obtained in the limit of small $l$, does not depend on $l$ as is justified 
by a negligibly weak dependence on $l$ of the order parameter $f_{l/2+0}$ on the end face of the external lead.

The proximity-modified current-phase relation $\tilde{\jmath}(\phi)$ is depicted at various $l$ in the right panel of 
Fig.\,\ref{fig:cur} for the normalized supercurrent $\tilde{\jmath}=j/j_{\text{dp}}$ and the interfaces with 
$g_\ell=g_\delta=0.1$. The supercurrent has been numerically evaluated, taking into account the phase incursion over the
central lead, based on \eqref{joscurr1} and the consistent solutions of GL equations for the system \eqref{Fb1}, 
\eqref{fint1}. A weakening of the regime of interchanging modes, taking place with decreasing $l$ at $l=1$ and $l=0.5$, 
is seen in the curves 1 and 2 for the set of parameters chosen.
 The pair breaking effects below $l\approx 0.36$ fully 
destroy the asymmetric states and a noticeable abrupt change of the supercurrent in the vicinities of $\phi_n=(2n+1)\pi$.
However, the curve 5 for $l=0.02$ within several percent coincides with the single junction sinusoidal current-phase 
dependence \eqref{joscurr2}. Anharmonic contributions to $\tilde{\jmath}(\phi)$ are seen in the curves 3 and 4. The 
dependence of the critical current on the central electrode's length $l$ in the double junction is shown in the left 
main panel of Fig.\,\ref{fig:cur}.

Thus, in the \mbox{SISIS} double Josephson junctions with closely spaced interfaces, the range of the internal phase 
differences is gradually reduced with decreasing the length $l$ of the central lead. The doubling of the current carrying
condensate states taking place at any given $\phi$, that occurs at $l\agt1$, is fully removed at very small $l$, when
the regime of interchanging modes is destroyed and the conventional single junction expression \eqref{joscurr2} for the 
current-phase relation is established.

The work has been carried out within the state task of ISSP RAS. 

\providecommand{\noopsort}[1]{}\providecommand{\singleletter}[1]{#1}%

\end{document}